\documentclass[aps,
prd,
preprint,
groupedaddress,showpacs,showkeys,12pt]{revtex4-1}
\usepackage{graphicx}
\usepackage{amsmath,amssymb,amsfonts} 
\usepackage{amscd}
\usepackage{amsthm}
\def\proof{{\bf Proof}.\ }

\newtheorem{Lemma}{Lemma}\newtheorem{Theorem}{Theorem}
\newtheorem{Definition}{Definition}
\newtheorem{Remark}{Remark}

\newfont{\goth}{eufm10}
\begin{document}
\title{On Foundations of Newtonian Mechanics}
\author{Al Cheremensky}%
\affiliation{IM--BAS, 4 Acad. G. Bonchev Str., Sofia--1113, Bulgaria\\
E--mail: {cheremensky@yahoo.com}}
\begin{abstract}\noindent
Being based on V. Konoplev's axiomatic approach to continuum mechanics, the paper broadens its frontiers in order to bring together continuum mechanics with classical mechanics in a new theory of mechanical systems. There are derived motion equations of `abstract' mechanical systems specified for  mass--points, multibody systems and continua: Newton--Euler equations, Lagrange equations of II kind and Navier--Stokes ones.
  
\noindent Quasi--linear constitutive equations are introduced in conformity with V. Konoplev's definition of stress and strain (rate) matrices. 

\end{abstract}
\medskip
 
\maketitle
 
 \section{Introduction} 

Classical mechanics is based on the axiom system introduced by I. Newton \cite{Newton}. 
In result of generalizations made by L. Euler it is also used to studying the kinematical and dynamical behavior of physical objects modeled as a rigid body or their aggregates. 

In the case of a parcel of air, water or rock consisting of a large number of particles, a corresponding discrete model, which can be constructed with the help of classical mechanics methods, would be hopelessly complicated. A different sort of models has been developed over the last three centuries to describe such physical systems. The model, called {\em continuous medium} or {\em continuum}, exploits the fact that in air, water and rock nearby particles behave similarly. The corresponding theory discounts the molecular structure of physical systems and regards matter as indefinitely divisible (here particles are characterized by their place volume and mass density). Thus the intent is to obtain a mathematical description of the macroscopic behavior of physical systems rather than to ascertain the ultimate physical basis of phenomena.
 
The analysis of the behavior of physical systems modeled as a continuum consists that we know as {\em continuum mechanics}.


A new architecture of mechanics is suggested in \cite{Konoplev1996,Konoplev1999} under the conditions that \vspace{-7pt}
\begin{enumerate}{\parskip -0.1cm \em
\item  there are no boxes or particles which can be rotated and deformed;
\item  there are no mass--points (points with zero volume and non--zero mass). }
\end{enumerate}

The first condition makes it essentially various w.r.t. conventional continuum mechanics while the second condition deepens the conflict between classical mechanics and that of continua \cite{Levi--Civita}:\vspace{-7pt}
\begin{quote}{\em 
`$\ldots $ the dynamics of a continuous system must clearly include as a
limiting case (corresponding to a medium of density everywhere zero except
in one very small region) the mechanics of a single material particle. This
at once shows that it is absolutely necessary that the postulates introduced
for the mechanics of a continuous system should be brought into harmony with
the modifications accepted above in the mechanics of the material particle'.}
\end{quote}
Following \cite{Konoplev1996,Konoplev1999,Truesdell} we aim to remove the conflict by bringing together the continuum mechanics of Konoplev and the classical one into a theory of mechanical systems. 

The new theory gives mathematical foundations to mechanics, which can be called {\em Newtonian} as 
it remains true to the principles of classical mechanics \cite{Newton,Truesdell} such as the absolute space and time, the concept of a mechanical system consisting of points in $3$--dimensional space as well as those of the mass additivity, actions--at--a--distance and differential laws of motion, Galileo's principle of relativity, {\em etc.} (it sounds curiously, but I. Newton has defined mass, as well as force, as `the measure of the same' -- see, {\em e.g.}, definitions I, II and VI in \cite{Newton} -- as though he has foreseen application of the measure theory to mechanics in 20th century \cite{Konoplev1996,Konoplev1999,Truesdell}).

The principle demand to a theory of mechanical systems is that `the problem of mechanics comes to describing motions
being in nature, namely, to their description in the most complete and
simple form' \cite{Kir}. Within the framework of this understanding the key concept of our theory is that a mechanical system is a set of points equipped  with some fields: the mass, force, velocity ones, {\em etc}.%

The theory is built on relatively simple, transparent ideas, some conventional notions are used, but sometimes their sense is radically changed. We try to give all of them on {\em tabula rasa} without using any background in the field of mechanics. That is why no prior knowledge of continuum mechanics or the classical is required. It does not mean that we have done all our best in order to avoid any mechanical reminiscences. However giving no comments or motivations, we are about to point out all technical details of the introduced constructions (for, as Goethe has told, `God is in the small things $\ldots $').


To demonstrate the new theory effectiveness we define the main classes of mechanical systems and deduce sufficiently many results known in the conventional mechanics: kinematics equations and Newton--Euler and Lagrange equations, stress--strain relations, {\em etc}.

We shall use the expression `see also' in the case where a given statement differs in details from that of cited works and thus it is formally absent in them.
 
\section{Paradigm of Newtonian mechanics}

The three laws of motion were first compiled by Sir Isaac Newton in his work `Mathematical Principles of Natural Philosophy', first published on July 5, 1687 (in Latin `Philosophiae Naturalis Principia Mathematica'). Newton used them to explain and investigate the motion of many physical objects and systems. For example, in the third volume of the text, Newton showed that these laws of motion, combined with his law of universal gravitation, explained Kepler's laws of planetary motion.

Newton's Laws hold only with respect to a certain set of frames of reference called Newtonian or inertial reference frames. Any reference frame that is in uniform motion with respect to an inertial frame is also an inertial frame, {\em i.e.}, Galilean invariance or the principle of Newtonian relativity.

Newton's first law is a restatement of the law of inertia. It apparently occurred to several different natural philosophers and scientists independently. Aristotle had the view that all objects have a natural place in the universe: that heavy objects like rocks wanted to be at rest on the Earth and that light objects like smoke wanted to be at rest in the sky and the stars wanted to remain in the heavens. He thought that a body was in its natural state when it was at rest, and for the body to move in a straight line at a constant speed an external agent was needed to continually propel it, otherwise it would stop moving.

The 17th century philosopher Ren\'{e} Descartes also formulated the law, although he did not perform any experiments to confirm it.

The first law (the inertia law), in less accurate form, has published still Galileo, and Newton gave credit to him. Galileo, however, realized that a force is necessary to change the velocity of a body, {\em i.e.}, acceleration, but no force is needed to maintain its velocity. This insight leads to Newton's First Law -- no force means no acceleration, and hence the body will maintain its velocity.

In the given interpretation mass, acceleration, momentum, and (most importantly) force are assumed to be externally defined quantities. This is the most common, but not the only interpretation: one can consider the laws to be a definition of these quantities.

Besides, Newton has definitively buried the representation which has taken roots from antique times that laws of motion of terrestrial and heavenly bodies are absolutely various. In its model of the world all Universe is subordinated to the uniform laws supposing the mathematical formulation.

Newton's laws were verified by experiment and observation for over 300 years, and they are excellent approximations at the scales and speeds of everyday life. Newton's laws of motion, together with his law of universal gravitation and the mathematical techniques of calculus, provided for the first time a unified quantitative explanation for a wide range of physical phenomena (here we do not discuss the relativistic mechanics).

Thus Newton's merit is the decision of two fundamental problems.
\vspace{-7pt}
\begin{enumerate}{\em 
\item	Creation of an axiomatic basis for mechanics which has actually passed this science in the category of strict mathematical theories.\vspace{-3pt}

\item	Creation of dynamics which connects behavior of a body with characteristics of external influences on it (forces).}\vspace{-3pt}\end{enumerate}

The Mathematical Principles of Natural Philosophy (Principia) is Newton's fundamental work in which it has formulated the law of universal gravitation and Newton's three laws being the base of the classical mechanics.

Newton defined main concepts -- mass, force, inertia (`congenital force of a matter'), quantity of motion, etc. The absoluteness of space and time which measure does not depend on position and speed of the observer are postulated. On the basis of these accurately certain concepts (they are a part of Newton's axiom system without which it is wrong.) three laws of Newtonian mechanics are formulated. While physicist Aristotle asserted that speed of a body depends on motive force, for the first time Newton made the essential amendment: not on speed, but on acceleration.

Newton's original Latin was translated quite closely by  Motte (1729). We shall give some definitions and the laws with the help of this translation.   

{\bf Definition I}. {\em 
The quantity of matter is the measure of the same, arising from its density and hulk conjunctly.}

{\bf Definition II}. {\em 
The quantity of motion is the measure of the same, arising from the velocity and quantity of matter conjunctly.} 

Note that the last notion is not quite clearly used at Descartes before. 

Now the quantity of motion is known as an impulse (of a body).

{\bf Definition III}. {\em 
The vis insita, or innate force of matter, is a power of resisting, by which every body, as much as in it lies, endeavours to persevere in its present stale, whether it be of rest, or of moving uniformly forward in a right line.

This force is ever proportional to the body whose force it is ; and differs nothing from the inactivity of the mass, but in our manner of conceiving it. A body, from the inactivity of matter, is not without difficulty put out of its state of rest or motion. Upon which account, this vis insita, may, by a most significant name, be called vis inertia, or force of inactivity. 
}

{\bf Definition IV}. {\em 
An impressed force is an action exerted upon a body, in order to change its state, either of rest, or of moving uniformly forward in a right line.

This force consists in the action only; and remains no longer in the 
body, when the action is over. For a body maintains every new state it 
acquires, by its vis inertice only. }

The vector nature of impressed forces is discovered in the parallelogram law (see Corollary II of the Principia - p. 84 in Motte's translation).

{\bf Definition V}. {\em 
A centripetal force is that by which bodies are drawn or impelled, or any way tend, towards a point as to a centre.}

{\bf Definition VI}. {\em 
The absolute quantity of a centripetal force is the measure of the same proportional to the efficacy of the cause that propagates it from the centre, through the spaces round about.}

{\bf Definition VII}. {\em 
The accelerative quantity of a centripetal force is the measure, of the same, proportional to the velocity which it generates in a given time.}

{\bf Definition VIII}. {\em 
The motive quantity of a centripetal force is the measure of the same, proportional to the motion which it generates in a given time.}

{\bf Law I}. {\em 
Every body perseveres in its state of rest, or of uniform motion in a right line, unless it is compelled to change that state by forces impressed thereon.}

Newton's first law postulates presence of such phenomenon, as inertia of bodies. Therefore it also is known as the inertia Law. Inertia is the phenomenon of preservation by a body of speed of motion (both on size, and in a direction), when on a body no forces or the vector sum of all operating forces operate (that is equally effective) is equal to zero. To change speed of motion, on a body it is necessary to work with some force. Naturally, the result of action of identical forces on size on various bodies will be various. Thus, say that bodies possess inertness. Inertness is a property of bodies to resist changing of their current condition. The inertness size is characterized by mass of a body.

It is necessary to notice that Galileo supposed free motion not only on a straight line, but also on a circle (it is visible, from astronomical reasons). Galileo has also formulated the major principle of the relativity, which Newton has not included in the axiomatic system because for mechanical processes this principle is a direct consequence of the equations of dynamics and (see the Principia):

{\bf Corollary V}. {\em The motions of bodies included in a given space are the same among themselves, whether that space is at rest, or moves uniformly forwards in a right line without any circular motion.} 

Newton considered space and time as absolute concepts, uniform for all Universe, and has obviously pointed out in the Principia. 

From the modern point of view, such formulation  of Law I  is unsatisfactory. First, `body' it is necessary to replace the term on `a material point' as the body of the final sizes for lack of external forces can make and a rotation  motion. Second, and this main thing, Newton in the work leant against existence of absolute motionless  frame , that is absolute space and time (the modern physics rejects this representation). On the other hand, in any (we will tell, rotating) frame the inertia law is incorrect. Therefore the Newton's formulation requires specifications.

{\bf Law II}. {\em 
The alteration of motion is ever proportional to the motive force impressed; and is made in the direction of the right line in which that force is impressed.}

The second law  states nothing about the innate force of matter as by definition IV only impressed forces are actions exerted upon a body, in order to change its state, either of rest, or of moving uniformly forward in a right line.


It is impossible to consider the first law as a special case of the second one where the motive (impressed) force is absent as  the former postulates existence of inertial frames while the latter is formulated already in such frames.

The vector nature of the second law addresses the geometrical relationship between the direction of the force and the manner in which the object's momentum changes. Before Newton, it had typically been assumed that a planet orbiting the sun would need a forward force to keep it moving. Newton showed instead that all that was needed was an inward attraction from the sun. Even many decades after the publication of the Principia, this counterintuitive idea was not universally accepted, and many scientists preferred Descartes' theory of vortices.

In a case when the mass of a material point time--invariant in due course, Newton's second law is usually formulated with use of the acceleration notion. 

Newton's second law -- the differential law of the motion describing interrelation between force applied at a material point and acceleration turning out from it of this point. Actually, Newton's second law enters mass as a measure of display of inertness of a body in a chosen inertial frame.

Some authors interpret the first law as defining what an inertial reference frame is; from this point of view, the second law only holds when the observation is made from an inertial reference frame, and therefore the first law cannot be proved as a special case of the second. Other authors do treat the first law as a corollary of the second. The explicit concept of an inertial frame of reference was not developed until long after Newton's death.

{\bf Law III}. {\em 
To every action there is always opposed an equal reaction: or the mutual actions of two bodies upon each other are always equal, and directed to contrary parts.}

The Third Law means that all forces are interactions between different bodies, and thus that there is no such thing as a unidirectional force or a force that acts on only one body. If body $A$ exerts a force on body $B$, body $B$ simultaneously exerts a force of the same magnitude on body $A$ -- both forces acting along the same line. Underline that these forces are enclosed to different bodies that is why at all are not compensated.

Newton's laws, strictly speaking, are fair only in inertial frames. If we fairly write down the equation of body motion in a non--inertial frame it will differ from Newton's second law by the form. However it is frequent, for consideration simplification, enter certain fictitious `force of inertia' and then the motion equations correspond in a kind very similar to the Newton's second law. In mathematical relation it is correct, but from the view point of mechanics it is impossible to consider the new fictitious force as something real, as result of some real interaction. Once again we will underline: `force of inertia' is only convenient convention in order that motion laws appear the same in inertial and non--inertial frames.

Not all motion equations  can be founded in the framework of Newtonian  mechanics described above. For example, we need the principle of constraint release in the case of non--free (constrained) bodies.

The sense or senses in which Newton used his terminology, and how he understood the second law and intended it to be understood, have been extensively discussed by historians of science, along with the relations between Newton's formulation and modern formulations. The modern interpretation of Newton's laws is given in \cite{Kokarew}.
 
With the definitions and laws given above one must connect the definitions and the laws of Newton with the concepts of classical mechanics  \cite{Einstein}  such as the concept of a body (medium) consisting of points in $3$--dimensional space as well as those of the mass additiveness, action--at--a--distance, Galilean (inertial) frame and Galileo's principle of relativity, principle of
Newton's determinacy, principle of release from constraints and so on. 
 
 It is important to note that all definitions and laws given above are stated without any mathematical expressions. Below we shall try to give them with the help of contemporary mathematics.
 
 \section{Main notions and principles of Newtonian mechanics}
    
In what follows we shall use {\em Galilean spacetime} \cite{Arnold} introduced as a quadruple ${\bf G} = \{{\bf A}_4, {\bf V}_4, g, \tau \}$ where\vspace{-7pt}
\begin{enumerate}{\parskip -.012cm \em 
\item  ${\bf V}_4$ is a 4--dimensional vector space,\vspace{-3pt}
\item  $\tau : {\bf V}_4 \rightarrow {\bf R}_1$ is a surjective linear map called the time map,\vspace{-3pt}
\item  $g=\langle \cdot,\cdot\rangle $ is an inner product on ${\rm  ker} \{\tau \}$ $(={\bf R}_3)$, and\vspace{-3pt}
\item  ${\bf A}_4$ is an affine normed space modeled on ${\bf V}_4$.} \vspace{-3pt}
\end{enumerate}

Introduce a point--wise spatial set ${\bf X}_3$ with the translation space ${\bf V}_3$ of $3$--dimensional (free) vectors and a parameterization $t\in {\bf R}_1 $ of the image of $\tau$ being in a point--wise time set ${\bf T}$ with the translation space ${\bf V}_{+}$ of $1$--dimensional (free) vectors having one and the same sense. For some parameterization $t\in {\mathbf R}_1 $ of $\mathbf T$ the differentiable map ${\mathbf R}_1 \rightarrow {\mathbf X}_3$ is called {\em motion}\index{motion}. \medskip 

{\bf  A. Screw space}

 It is considered as conventional \cite{Dimentberg} that the screw calculus is not adapted for the description of continuous media, and `$ \ldots $ being very attractive representation of a system of forces and rigid body motions with the help motors and screws, nevertheless it has no essential practical value $ \ldots $' \cite{Sommerfeld}. As a result in mechanics there is mainly absent the fundamental understanding (concept) that the interaction between mechanical systems is described with the help of screws. 

The using of the screw concept is the key for the theory of mechanical systems (including, a continuum, a mass point and a rigid body) which is below constructed.
 
  Let us define two fields ${\mathfrak R }$ and ${\mathfrak M }$ of vectors attached to points of ${\bf X}_3$ such, that for any two points $a$ and $b\in {\bf X}_3$ with $\vec{r}_a$, $\vec{r}_b\in {\mathfrak R }$ and $\vec{\mu}_a$, $\vec{\mu}_b\in {\mathfrak M }$ there is the following relation 
\begin{equation}\vec{r}_a=\vec{r}_b\stackrel{\rm { def}}{=} \vec{r}, \quad \vec{\mu}_a=\vec{\mu}_b+\overrightarrow{ab}\times \vec{r } 
\label{ 1}\end{equation}
where $\times $ means vector product.

\begin{Definition}{\rm \cite{Berthelot}} 
The set $\{{\mathfrak R }, {\mathfrak M }\}$ is called {\em screw} while $\{\vec{r}, \vec{\mu}_a\}$ is {\em element of reduction} of the screw  at a point $a\in {\bf X}_3$. The vectors $\vec{r}$ and $\vec{\mu}_a$ are called {\em main vector} (resultant) and {\em total moment} of the screw (at the point $a$), respectively (the screw moment is not that of the vector $\vec{r}$ as we do not connect the point $a$ with a point in ${\bf X}_3$).
\end{Definition}
We do not support the idea to use the name `torser' from the French word `torseur' instead of `screw'
 \cite{Berthelot}.

A screw with the property $\vec r \times \vec{\mu}_a=0$ (for all points $a\in {\bf X}_3$) is called {\em slider} \cite{Berthelot}. We say that a slider is axial at some point $a$, if $\vec{\mu}_a=0$ (we have at least one such point). It is useful to note that the sum of axial sliders is the axial slider, too.

A screw with the property $\vec r =0$ is called {\em couple}, from (\ref{ 1}) follows that $\vec{\mu}_a=\vec{\mu}_b\stackrel{\mathrm{def}}{=}\vec{\mu}$ for all points $a$ and $b\in {\bf X}_3$ \cite{Berthelot}. 

\begin{Definition}We shall call {\em wrench} and {\em twist} the following operator forms, respectively: 
\[\pi_a ^{wr}=\begin{pmatrix} \vec {r } \\ \vec {\mu}_a
\end{pmatrix}, \quad \pi_a ^{tw}=\begin{pmatrix} \vec {\mu}_a \\ \vec {r } 
\end{pmatrix}
\]
defined at the point $a\in {\bf X}_3$.
\end{Definition}
We may also define them as  objects with the following properties:
\[\pi_a ^{wr}=\begin{bmatrix} I &\ O \\ \overrightarrow{ab}\times 
&\ I \end{bmatrix} \pi_b ^{wr},\quad \pi_a ^{tw}= \begin{bmatrix} I\ & \overrightarrow{ab}\times \\ O \ 
& I \end{bmatrix} \pi_b ^{tw}\]
where $I$ and $O$ are unit and zero matrices.

Henceforth we shall briefly say {\em slider }\index{Slider} a wrench with the slider property and note it as $l^r$.

Define the triple of orthogonal unit vectors 
$\vec{\bf e}^{\hspace{.05cm } a}=\{
\vec e_1^{\hspace{.1cm } a}, 
\vec e_2^{\hspace{.1cm } a}, 
\vec e_3^{\hspace{.1cm } a}\}$ in the $3$--dimensional space ${\bf V}_3$.  Let us introduce  $6$ wrenches (twists) such that  at the point $a\in {\bf X}_3$ their elements are defined as follows\vspace{3pt}
\[
{\mathfrak e}_1^{ a}=\begin{pmatrix} \vec e_1^{\hspace{.05cm } a} \\ \vec o
\end{pmatrix},\  
{\mathfrak e}_2^{ a}=\begin{pmatrix} \vec e_2^{\hspace{.05cm } a} \\ \vec o
\end{pmatrix},\ 
{\mathfrak e}_3^{ a}=\begin{pmatrix} \vec e_3^{\hspace{.05cm } a} \\ \vec o
\end{pmatrix},\ 
{\mathfrak e}_4^{a}=\begin{pmatrix} \vec o \\ \vec e_1^{\hspace{.05cm } a}
\end{pmatrix},\  
{\mathfrak e}_5^{ a}=\begin{pmatrix} \vec o \\ \vec e_2^{\hspace{.05cm } a}
\end{pmatrix},\  
{\mathfrak e}_6^{ a}=\begin{pmatrix} \vec o \\ \vec e_3^{\hspace{.05cm } a}
\end{pmatrix}
\]
 where $\vec  o\in {\bf V}_3$ is the null vector.
 
 As any screw is defined in the unique way by its element of reduction at some point, these $6$ wrenches (twists) generate the basis ${\hat {\mathfrak e}}^{\hspace{.05cm } a}$ of the screw space, the first triple of the wrenches being axial sliders and the second one being couples. As elements of the spaces ${\bf V}_3$ and ${\bf R}_3$ are called {\em vectors}, we may use the names {\em wrench}, {\em twist}   and {\em slider} for  coordinate columns of  elements of any wrench, twist or slider in the basis ${\hat {\mathfrak e}}^{\hspace{.05cm } a}$ at the  point $a\in {\bf X}_3$.
 
A screw can be resolved in a sum of a slider and a couple if it is neither slider nor couple \cite{Berthelot}. This resolution is not unique. A couple can be represented as sum of two sliders. That is why any screw (as a vector in the screw space ${\bf S}$) is a slider sum, too. This fact is used  \cite{Konoplev1996} in order to define a slider as the primary notion of screw theory.
\begin{Remark}For a given system of line vectors there exists a point $a\in {\bf X}_3$ such that its main vector $\overrightarrow{r}$ and total moment  $\overrightarrow{\mu}_{\hspace{ -.08cm} a}=0$ are such that  $\overrightarrow{r}\times \overrightarrow{\mu}_{\hspace{ -.08cm} a}=0$. In this case the 
system is called {\em screw} being the set of the following elements \cite{Yakovenko}
\vspace{-11pt}\begin{quote}{\parskip -.4cm
\item \thinspace \thinspace \hskip -.7cm --- \thinspace \thinspace 
the straight line {\em (screw axis)} passing through the point $a$;
\item \thinspace \thinspace \hskip -.7cm --- \thinspace \thinspace 
the main vector $\overrightarrow{r}$ giving the screw axis sense;
\item \thinspace \thinspace \hskip -.7cm --- \thinspace \thinspace 
the moment vector $\overrightarrow{\mu}_{\hspace{ -.08cm} a}$ (being collinear to $\overrightarrow{r}$).}
\end{quote}\vspace{-7pt}
One says that the point $a$ is that of the screw reduction and this screw depicts a screw motion.

This definition leads to no matrix tools of screw calculus which may simplify the reduction of line vectors to  the simplest equivalent system  \cite{Yakovenko}.
\end{Remark} \newpage

{\bf  B. Main measures of Newtonian mechanics {\rm (see also 
{\cite{Konoplev1996,Konoplev1999}})}}

Let us define the Lebesgue measure $\mu _t$ on $\sigma $--algebra of subsets in ${\bf T}$  while on $\sigma $--algebra of subsets in ${\bf X}_3$  there be so called  {\em determinative} time--invariant measure
\[
\mu_{_{LS}} (A)=\mu _{ac}(A)+\mu _{pp}(A) 
\]
where $\mu _{ac}(A)$ is the absolutely continuous component w.r.t. Lebesgue
measure $\mu _3$ and $\mu _{pp}(A)$ is the pure point (discrete) component
presented as $\mu _{pp}(A)=\sum_k\mu _{pp}({x }_k)$ for points in an
arbitrary subset $A\in { \sigma }_3$ such that $\mu _{pp}({x }_k)\neq 0$.
These points are called {\em pure}, the others being called {\em continuous} \cite{Reed}. We assume $\mu _{ac}$ to be Lebesgue measure $\mu _3$.

We shall further use the measure $\mu_{_{LS}}$ for definition of points with mass, but without volume, and bodies with volumes, but without masses and forces exerting on them.

\begin{Definition}  Let $A\in { \sigma }_3$, then the measure  $m(A): A\rightarrow {\bf R}_1$ is called {\em mass} {\em (measure of inertia)}. 
\end{Definition}
Due to the Radon--Nikodym theorem \cite{Reed} we may specify $m(A)$ as Lebesgue--Stieltjes integral 
\[
m(A)=\int \hspace{-.05cm} \chi _{_{A}}\rho _x\mu_{_{LS}} (dx)
\]
with a $\mu_{_{LS}} $--integrable (mass) density $\rho _x$ w.r.t. the measure $\mu_{_{LS}}(dx)$ (here $\chi _{_{A}}$ is the {characteristic function} of ${A}$). The density can be time--varying.

The set ${\bf X}_3^c \subset {\bf X}_3$ is called {\em set of concentration} of the measure $m$ on ${\bf X}_3$ if $m(B)=0$ for everyone $\mu_{_{LS}}-$measured set $B \subset {\bf X}_3\setminus {\bf X}_3^c$.

We shall use the notion of {\em signed measure} \cite{Evans} being a generalization of the concept of measure by allowing it to have negative values. Some authors call it {\em charge}, by analogy with electric charge, which is a familiar distribution that takes on positive and negative values.

Let $ \eta (\cdot )\in {\bf V}_3$ be a function on $\sigma _3$ whose
components (in some basis) are signed measures. Then: \vspace{-7pt}
\begin{enumerate}\parskip -0.15cm 
\item  the function $ \eta (\cdot )$ is called {\em vector signed measure} on $\sigma _3$; 
\item  a function $\zeta (\cdot ,\cdot )\in {\bf V}_3$, defined on $\sigma _3\times
\sigma _3$ and being a vector signed measure by each of arguments, is
called {\em vector signed bi--measure};
\item  the vector signed bi--measure $\zeta (\cdot ,\cdot )$ is called
skew if 
$\zeta (A,B)=-\zeta (B,A)$ for any $A$ and $B\in \sigma _3$.
\end{enumerate}
\begin{Definition}  Given $A$ and $B\in { \sigma }_3$,  the skew vector signed bi--measure ${\mathcal F}(A,B)\hskip -.1cm : (A,B)\rightarrow {\bf S}$ is called {\em measure of action} of $B$ on $A$.  
\end{Definition}
\begin{Remark}It is important to point out that C. Truesdell defines mainly the measure of action as $3-$dimensional vector \cite{Truesdell}.
\end{Remark}
We specify ${\mathcal F}(A,B)$ as Lebesgue--Stieltjes integral 
\[{\mathcal F}(A,B)=\int \hspace{-.05cm} \chi _{_A} l^{\phi(x,B)} \mu_{_{LS}} (dx)=
\int \hspace{-.05cm} \chi _{_B} l^{\psi(y, A)}\mu_{_{LS}} (dy)\]
where elements of $\mu_{_{LS}} $--integrable slider functions $l^{\phi(x,B)}$ and $l^{\psi(y, A)}$ are axial at $x\in A$ and $y\in B$,
respectively (they can be represented as corresponding Lebesgue--Stieltjes integrals with densities being axial sliders).

The set $A^e={\bf X}_3\setminus 
A$ is called {\em environment} of $A$. 
It is clear that 
\begin{equation}
{\mathcal F}(A,A^e)=
{\mathcal F}(A,A^e+A)\stackrel{\rm { def}}{=}\int \hspace{-.05cm} \chi _{_A} l^{\phi(x,A^e+A)}
\mu_{_{LS}} (dx)=\int \hspace{-.05cm} \chi _{_A} l^{\phi(x,x^e)}\mu_{_{LS}} (dx)
\label{env}
\end{equation}
 We shall assume that $l^{\phi(x, x^e)}\equiv 0$ on the set ${\bf X}_3\setminus {\bf X}_3^c$.
\begin{Definition}  The slider function $l^{\phi(x,x^e)}$ is called {\em intensity} of the action of $x^e$ upon $x\in {\bf X}_3^c$.
\end{Definition}
\begin{Remark}  The intensity can also depend on the motion prehistory. \end{Remark}
Exemplify the introduced notion. Let the skew bi--measure ${\mathcal G}(A,B)\in {\bf V}_3$ be  such that 
\[
{\mathcal G}(A,A^e) =\int \hspace{-.05cm} \chi _{_A} l^{g(x, x^e)} \rho_x \mu_{_{LS}}(dx),\quad \vec g(x, x^e)=\gamma 
\int \hspace{-.05cm} \chi _{_{x^e}} 
\overrightarrow{ (x-y)}
\frac{\rho_y \mu_{_{LS}}(dy)}{\Vert
\overrightarrow{ (x-y)}\Vert ^3}
\]
where $\gamma $ is a positive (gravitational) constant, elements of the $\mu_{_{LS}} $--integrable slider function $l^{g(x, x^e)}$ are axial at $x\in {\bf X}^c_3$.
\begin{Definition}{\rm \cite{Konoplev1999}}
The slider function $\rho_x l^{g(x, x^e)}$ is called 
{\em intensity of gravitating action} of $x^e$ upon $x\in {\bf X}^c_3$.
\end{Definition} 

 {\bf  C. Fundamental principles of dynamics} 

Let ${\sigma }_t$ be Borel $\sigma $--algebra of subsets in ${\bf T}$ while ${ \sigma }_3$ is Borel $\sigma $--algebra of subsets in ${\bf X}_3$.

Let us fix some parameterization $t\in {\bf R}_1 $ of $\bf T$, then the differentiable bijection: ${\bf X}_3 \rightarrow {\bf X}_t\subset {\bf R}_3$ is called {\em motion}, $t$ is a time instant. 

For any point $x\in {\bf X}_3$ the motion defines the point $x(t)\in {\bf X}_t$. Introduce the radius--vector $\vec {r}_{x}(t)=\overrightarrow{(O_0,x(t))}$ called {\em position} of $x(t)\in {\bf X}_t$ and the vector $\vec {v}_{x}=\vec {r}_{x}(t)^{\centerdot}
$  called its {\em velocity} w.r.t. $O_0$. Thus we equip the  set 
${\bf X}_3$ with the fields of positions, velocities and the measures of mechanics. 

 Let the slider  $l^{v_{x}}$ be axial at $x\in {\bf X}_3$.
\medskip

\noindent{\bf Second Newton's law} (see also \cite{Konoplev1999,27}). {\em There exist a Cartesian frame $\mathcal E_0$ with the origin $O_0$ and a parameterization $t\in {\mathbf R}_1 $ of $\mathbf T$ such that  motion of a point $x\in {A}^c$ 
  is described by the following equations in the slider form \vspace{-7pt}
\begin{enumerate}{\parskip -0.15cm 
\item  if the  point $x$ is continuous
\begin{equation}\rho _x
(l_x^{v_{x},0})^{\centerdot}=l_x^{\phi(x,x^e),0}\label{ 2n}
\end{equation}
\item  if the  point $x$ is  pure 
\begin{equation}m_x
(l_x^{v_{x},0})^{\centerdot}=\mu _{pp}(x) l_x^{\phi(x,x^e),0}\label{ 3n}
\end{equation}
where $m_x=\rho _x \mu _{pp}(x)$ is {\em mass} of the pure point (coordinate representations of vectors in $\mathcal E_0$ are marked with the superscript ${}^0$ while the subscript ${}_0$ means that the slider moment is computed w.r.t. the point $x$; to honor Newton,   we use the superscript ${}^\centerdot$ for derivatives by $t$).}
\end{enumerate}}
Henceforth we call the parameterization and the frame ${\cal E}_0$ {\em Galilean} (this formulation of second Newton's law is connected with first one and isolated systems nohow).

\begin{Remark}  In the case of time--varying densities of inertia (masses) relations (\ref{ 2n})--(\ref{ 3n}) are invariant w.r.t. {\em Galilean group} \cite{Konoplev1996,Konoplev1999} while the traditional form of second Newton's law \cite{Newton} does not. In such case  relations (\ref{ 2n})--(\ref{ 3n}) include in themselves slider functions of so called reactivity (see the well--known equation of Mescerskii). \end{Remark}

\begin{Definition} The set ${\bf G}$ (${\bf X}_3, {\bf T}$), $(\sigma _3, \sigma _t)$ and $(\mu_{_{LS}}, m, {\mathcal F})$ answering the   second Newton's law and the principles of causality, determinacy and relativity \cite{Arnold} are called {\em Universe of Newtonian mechanics}, elements of $\sigma -$algebra $\sigma _3$ being called {\em mechanical  systems} \cite{Konoplev1999}.
\end{Definition}
In the given definition {(see also  \cite{Konoplev1996,Konoplev1999})}, similarly to that of {\em probability space} \cite{Kolmogorov}, Universes of mechanics are separately specified for every mechanical problem.

From relations (\ref{ 2n})--(\ref{ 3n}) follows that 
motion of a point $x\in {A}^c$ 
  is described in the following slider form
\begin{enumerate}{\parskip -0.15cm \em
\item if the  point $x$ is continuous
\begin{equation}\rho _x
(l_0^{v_{x},0})^{\centerdot}=l_0^{\phi(x,x^e),0}\label{ 2}
\end{equation}
\item if the  point $x$ is  pure 
\begin{equation}m_x
(l_0^{v_{x},0})^{\centerdot}=\mu _{pp}(x) l_0^{\phi(x,x^e),0}\label{ 3}
\end{equation}}
\end{enumerate}
It is easy to see that all slider forms of motion lead to the following vector relations
\begin{enumerate}{\parskip -0.15cm \em
\item  if the  point $x$ is continuous
\begin{equation}\rho _x
v_x^{0\centerdot}=\mu _{pp}(x)\phi^0 (x,x^e)\label{ 2p}
\end{equation}
\item  if the  point $x$ is pure 
\begin{equation}m_x
v_x^{0\centerdot}=\mu _{pp}(x)\phi^0 (x,x^e)\label{ 3p}
\end{equation}}
\end{enumerate}

In the motion equations the intensities are defined nohow, and any action intensity pictures some mechanical system \cite{Arnold} in depending on its `constitution'. Sometimes  some part of the intensities is implicitly given, while another one must be defined from  the restriction or constraint imposed on a point, its velocity and, perhaps, derivative of the velocity, beforehand set, {\em i.e.}, not dependent on the law of point  motion. In this case a point which motion is in agreement with constraints is called {\em 
constrained}.\medskip


\noindent {\bf Example}.  Let the vector $\vec \phi(x,x^e)$ describe the action of $x^e$ on $x$ when constraints are absent and the  constraints be given by the equation
 $\sigma     (r_x^0, v_x^{0},t)=0$ 
where $\sigma     $ is a differentiable vector--function of the instant $t$, the position $r_x^0$ and the velocity $v_x^{0}$.  After differentiating ${\sigma    }(r_x^0,v_x^0,t)=0$ we have 
\[
\frac{\partial \sigma    }{\partial r_x^{0,T}}
v_x^{0}+
\frac{\partial \sigma    }{\partial v_x^{0,T}}
v_x^{0\centerdot}+\frac{\partial \sigma    }{\partial 
t}=0
\]and 
\[v_x^{0\centerdot}=-(\frac{\partial \sigma    }{\partial v_x^{0,T}})^T 
[\frac{\partial \sigma    }{\partial v_x^{0,T}}(\frac{\partial \sigma    }{\partial v_x^{0,T}})^T]^{-1}(
\frac{\partial \sigma    }{\partial r_x^{0,T}}
v_x^{0}+\frac{\partial \sigma    }{\partial 
t})
\]
if the above inverse exists. 

Hence there exists such
 slider function $l^{c(x,x^e)}$ (axial at $x\in A^c$) that  
\begin{equation}\rho _x
(l_0^{v_{x},0})^{\centerdot}=l_0^{\phi(x,x^e),0}+l_0^{c(x,x^e),0}\label{ 13a}
\end{equation} if  the  point $x$ is continuous or 
\begin{equation}m_x
(l_0^{v_{x},0})^{\centerdot}=\mu _{pp}(x) [l_0^{\phi(x,x^e),0}+l_0^{c(x,x^e),0}]\label{ 13b}
\end{equation}
if the  point $x$ is  pure.

In this way we may introduce the following \medskip

\noindent{\bf Principle of constraint release}. {\em Motion of any constrained point $x\in A^c$   is described by  equations (\ref{ 13a}) and (\ref{ 13b}) (in the Galilean frame ${\cal E}_0$)  
with some $\mu_{_{LS}} $--integrable slider function $l^{c(x,x^e)}$ called 
{\em intensity of constraint action}  upon $x\in A^c$}.

The principle of constraint release demarcates two categories of actions, namely, active and
passive ones: it says that an active (motive) action creates motion while a passive
one only puts obstacles in this motion. If we remove constraints then only active actions are kept.
\begin{Remark} One must not suppose that the principle of release from constraints and
that of D'Alembert eliminate the difference in the nature of active forces and passive
ones (constraint actions and forces of inertia). It is only for the sake of convenience that we use
these principles: only forces the resultant of which is $f$  exert on a point (body) \cite{Ban}.\end{Remark}

\section{A mass--point}
Consider a set $A\in { \sigma }_3$ consisting of a unique pure point of the measure $\mu_{_{LS}} $ as a free {\em mass--point}. In this case from relation (\ref{ 3}) follows the well--known {\em second Newton's law} (in the Galilean frame ${\cal E}_0$):%
\begin{equation}
m_x v_{x}^{0\centerdot}=f^0_x
\label{ 11}
\end{equation}
where $\vec f_x=\mu _{pp}(x)\vec \phi (x,x^e) $ is called {\em force} exerting on the mass--point $x$. 

Note that in the case where the mass is time--varying the force includes in itself that of reactivity (see the well--known equation of Mescerskii). 

Motion of a mass--point can be constrained. Let us give the description of constrained motion.

Variety of constraints contains so called ideal and non--ideal ones. Ideal constraints generate
constraint actions having the direction and sense of the normal to the corresponding
manifold. We shall assume that constraints are ideal ({\em Axiom of
ideal constraints}), scleronomic and holonomic.

Ideal holonomic and scleronomic constraints force the point under consideration to move along with a certain manifold $\sigma    (r_x^0)=0$ having lower dimension than its
configuration space. Let this manifold can be parameterized with some vector ${q}$. The vector ${q}$ is called {\em generalized} one, its first  derivative being
called {\em generalized velocity} ${q}^{\centerdot }$.

For any point $x$ of the manifold we have 
\[r_x^0= \eta ({q})
\]
If the  columns of the following matrix \cite{Vel}
\[\tau = \frac{\partial \eta}{\partial {q}^T}
\]
are linearly independent, they form a basis of the linear space ${\bf T}={\bf T}({q})$ being tangent to the manifold at a point ${q}$.

If the  columns of the following matrix \cite{Vel}
\[ \nu=\frac{\partial \sigma     }{\partial r_x^{0,T}}
\] are linearly independent, they form a basis of the linear space ${\bf N}={\bf N}(q)$ being orthogonal to the manifold at a point $q$.

Thus ${\bf R}_3={\bf T}\times {\bf N}$ with the basis $[{\bf \tau,\nu}]$.
It is easy to see that
\[P_{\tau } {\bf R}_3={\bf T}, \quad P_{\nu }{\bf R}_3= {\bf N}
\]
where $P_{\tau }=(\tau ^T\tau )^{-1}\tau ^T$ and $P_{\nu }=\nu(\nu^T\nu)^{-1}\nu^T$ are projections.

It is obvious that $r_x^{0\centerdot }=\tau{q}^{\centerdot }$ and $r_x^{0\centerdot \centerdot }=\tau{q}^{\centerdot \centerdot}+
(\frac{\partial \tau}{\partial {q}^T}{q}^{\centerdot })
{q}^{\centerdot }$ (here the matrix $\frac{\partial \tau}{\partial {q}^T}{q}^{\centerdot }$ is square).

As the constraint is supposed to be ideal and therefore ${ P }_{\tau}{r_x^0}=0$, from (\ref{ 13b}) follows 
\begin{equation}m_x[{q}^{\centerdot \centerdot}+{ P }_{\tau}
(\frac{\partial \tau}{\partial {q}^T}{q}^{\centerdot })
{q}^{\centerdot }]=P_{\tau }{f^0_x}
\label{ 25}
\end{equation}

Applying the projection $P_{\nu }$ to (\ref{ 13b}) we  define the following equation 
\[m_xP_{\nu }(\frac{\partial \tau}{\partial {q}^T}{q}^{\centerdot })
{q}^{\centerdot }=P_{\nu }(f^0_x+c^0_x)=P_{\nu }f^0_x+c^0_x
\]

From this relation follows that
\[c^0_x=
m_xP_{\nu }(\frac{\partial \tau}{\partial {q}^T}{q}^{\centerdot })
{q}^{\centerdot }-P_{\nu }f^0_x
\]
{\em i.e.}, the constraint force is not a function of time, but it depends on the generalizing coordinates and velocities as well as on the active force. 

It is clear that the theory above can be applied to mass--point systems.

\section{Rigid bodies}
\begin{Definition} {\rm (see also \cite{Vilke})}. A bounded closed set $A\in {\sigma }_3$ is called {\em rigid body} if 
\vspace{-7pt}
\begin{enumerate}{
\item  constraints applied on its points keep distances between them not changing with time;\vspace{-7pt}
\item  the constraints are ideal.}
\end{enumerate}\vspace{-11pt}
\end{Definition}
A rigid body may contain continuous and pure points.
\begin{Remark}  In elementary manuals of mechanics, transition from a mass  point to a body as a point system is made somehow imperceptibly; constraint forces are not mentioned at all, and instead of a lawful exception there is an illegal, silent exclusion of these forces. They remain ordinarily without any attention and even without a mention, as if they did not exist at all.
\end{Remark}

 
{\bf Multiplicative groups of motions}. In  ${\bf V}_3$ let us have 3 triples of orthogonal unit vectors 
$\vec{\bf e}^{\hspace{.05cm } 0}=\{
\vec e_1^{\hspace{.1cm } 0}, 
\vec e_2^{\hspace{.1cm } 0}, 
\vec e_3^{\hspace{.1cm } 0}\}$, $
\vec {\bf e}^{\hspace{.05cm} {p}}=\{
\vec e_1^{\hspace{.1cm } {p}}, 
\vec e_2^{\hspace{.1cm } {p}}, 
\vec e_3^{\hspace{.1cm } {p}}\}$ and $
\vec {\bf e}^{\hspace{.05cm } {k}}=\{
\vec e_1^{\hspace{.1cm }{k}}, 
\vec e_2^{\hspace{.1cm }{k}}$, 
$\vec e_3^{\hspace{.1cm }{k}}\}$, where ${p}$ and ${k}$ are naturals (here we may use $p=0$, $p=k-1$ and $p=k$). 

Assume that $\vec{\bf e}^{\hspace{.05cm } 0}$ is chosen as the main basis and the space ${\bf R}_3$ has so called canonical 
 basis ${\bf e}^0$, consisting of 
$e_1^0={\rm col}\{ 1, 0,0 \}$ and $e_2^0={\rm col}\{0, 1,0 \} $, $e_3^0= {\rm col}\{0, 0,1 \}$. The other bases can be movable. With their help let us define the frames ${\cal E}_0={\cal E}(O_0,{\bf e}^{ 0})$, ${\cal E}_p={\cal E}(O_p,{\bf e}^{p})$ and ${\cal E}_k={\cal E}(O_k,{\bf e}^{k})$ with the origins $O_0$, $O_p$ and $O_k$. 

Define rotation matrices $C_{0, {p}}$ and $C_{{p},{k}}$ such that $C_{0, {p}}C_{{p},{k}}= C_{0, {k}}$ and for any free vector  $\vec{\lambda}$ there are the following relations
\[\lambda^0= C_{0,{p}}\lambda^{p}, \quad \lambda^{p}= C_{{p},{k}}\lambda^{k}\vspace{-7pt}\] 
where $\lambda^0$, $\lambda^{p} $ and $\lambda^{k} \in {\bf R}_3$ are the coordinate columns of the vector $\vec{\lambda}$ in the bases $\vec{\bf e}^{\hspace{.05cm } 0}$, 
$\vec{\bf e}^{\hspace{.05cm } {p}}$ and $\vec{\bf e}^{\hspace{.05cm } {k}}$. Hence we have also $\lambda^0= C_{0,{k}}\lambda^{k}$.

Let $x$ be an arbitrary point fixed in ${\cal E}_p$. Introduce the radius--vectors $\vec {r}_0$ and $\vec {r}_p\in {\mathbf V}_3$ of
the point $x$ w.r.t. the origins $O_0$ and $O_p$, respectively. Define $\vec d_{0,p}=\vec {r}_0-\vec {r}_p$. Then we may 
represent the relation $\vec {r}_{0}=\vec {d}_{0,p}+\vec {r}_{p}\in {\bf V}_3$ in ${\cal E}_0$ as $r^0_0=d^0_{0,p}+C_{0,p}r_p^p$. As $r_p^p$ is time--constant, with differentiating the last relation we have 
$v^0_x=v^0_{0,p}+C_{0,p}^{\centerdot  }r_p^p$ where $v^0_x=r_0^{0\centerdot }$ and $v^0_{0,p}=d_{0,p}^{0\centerdot }$ are velocities of $x$ and $O_p$ w.r.t. $O_0$ in the frame ${\cal E}_0$, respectively. Hence 
\[
v_x^p=v_{0,p}^p+C_{p,0}C_{0,p}^{\centerdot }r_p^p
\vspace{-11pt} 
\]

For any vector $f=\begin{pmatrix}
f _1 \\ 
f _2 \\ 
f _3 
\end{pmatrix}\in {\bf R}_3$ introduce the cross product matrix\vspace{-25pt}   \begin{equation}
 f^{\times}\stackrel{\mathrm{def}}{=} \hspace{-0.12cm}\begin{bmatrix}
0 & -f_3 & f _2 \\ 
f_3 & 0 & -f _1 \\ 
-f_2 & f _1 & 0
\end{bmatrix}\label{ 2g}
\end{equation}

Let us define (in ${\cal E}_{p}$) \cite{Lurie}:
\vspace{-11pt}\begin{quote}{\parskip -.24cm
\item \thinspace \thinspace \hskip -.7cm --- \thinspace \thinspace 
the coordinate column $d_{{0},{p}}^{p}\in {\bf R}_3$ of  the {\em translation} vector of ${\cal E}_{p}$ w.r.t. ${\cal E}_{0}$;
\item \thinspace \thinspace \hskip -.7cm --- \thinspace \thinspace 
the coordinate column  $v_{{0},{p}}^{p}\in {\bf R}_3$ being known as {\em quasi--velocity} of the translation of ${\cal E}_{p}$ w.r.t. ${\cal E}_{0}$;
\item \thinspace \thinspace \hskip -.7cm --- \thinspace \thinspace 
the cross product matrix $\omega _{{0},{p}}^{{p}\times} \stackrel{\rm { def}}{=} C_{p,0} C_{{0},{p}}^{\centerdot}$ where the triple $\omega _{{0},{p}}^{p}\in {\bf R}_3$ is known as {\em angular quasi--velocity} of rotation of ${\cal E}_{p}$ w.r.t. ${\cal E}_{0}$ and is the eigenvector of $C_{{0},{p}}$, answered with the eigenvalue $1$;
\item \thinspace \thinspace \hskip -.7cm --- \thinspace \thinspace 
{\em quasi--velocity}
$V_{{0},{p}}^{p}={\rm col}\{v _{{0},{p}}^{{p}},\omega _{{0},{p}}^{{p}}\} \in {\bf R}_6$  of motion of ${\cal E}_{p}$ w.r.t. ${\cal E}_{0}$.}\vspace{-3pt}
\end{quote}
These algebraic quantities are answered with geometrical ones, {\em e.g.}, vectors of the translation velocity $ \vec v_{{0},{p}}\in {\bf V}_3$ and the instantaneous angular 
velocity $ \vec\omega _{{0},{p}}\in {\bf V}_3$ which are defined with the help of the basis $\vec{\bf e}^{\hspace{.05cm } {p}}$. We may use them in order to define the twist (kinematical screw) $V$ with the element $V_{{0},{p}}={\rm col}\{\vec v _{{0},{p}},\vec\omega _{{0},{p}}\}$ at the point $O_p$.

In the kinematics the angular velocity $ \vec\omega _{{0},{p}}$ defines the rotation axis of ${\cal E}_{p}$.

Introduce following matrices
\begin{equation}
{ T}^0_{0,p}=
\begin{bmatrix}
I &\ O\\ 
 \hspace{ .15cm}d_{0,p}^{0\times } &\ \hspace{ .15cm}I
\end{bmatrix}, \quad 
{C}_{0,p}^\otimes 
=  
\begin{bmatrix}
C_{0,p} & O \\ 
O & C_{0,p}
\end{bmatrix}
\label{ 3g}
\end{equation}
\begin{Theorem} {\rm \cite{Konoplev1999}} Let us have a wrench $\pi ^{wr}$. Then $\pi_0 ^{wr, 0}= {L}^{wr}_{0,p}\pi_p^{wr ,p}$ where  
$\pi_0 ^{wr}$ and $\pi_p ^{wr}$ are its elements at the points $O_0$  and $O_p$, the matrix ${ L}^{wr}_{0,p}$ has the representation 
\begin{equation}
	{ L}^{wr}_{0,p}={ T}^0_{0,p} {C}_{0,p}^\otimes={C}_{0,p}^\otimes { T}^p_{0,p}\label{66}
\end{equation}
 and belongs to the multiplicative group ${\mathcal L}^{wr}({\mathcal R}, 6)$ such that   
\begin{equation} 
{ L}^{{wr}\centerdot }_{0,p}={ L}^{wr}_{0,p}{\it \Phi}_{0,p}^{wr}, \quad {\it \Phi}_{0,p}^{wr}=
\begin{bmatrix}
 \omega_{0,p}^{p\times} & O \\ 
 v_{0,p}^{p\times} & \omega_{0,p}^{p\times}
\end{bmatrix}
\label{ 4}
\end{equation}
\end{Theorem}
\proof The representation of ${ L}^{wr}_{0,p}$ follows directly from the screw  definition.

Relation (\ref{ 4}) is true as from (\ref{ 3g}) follows that 
 ${ L}^{{wr}\centerdot}_{0,p}={ T}^{0\centerdot}_{0,p}C_{0,p}^\otimes +{ T}%
^0_{0,p}{C}^{\otimes \centerdot}_{0,p}={ T}^0_{0,p}
{C}_{0,p}^\otimes ({ C}_{p,0}^\otimes { C}^{\otimes \centerdot}_{0,p}$ $+{ C}_{p,0}^\otimes { T}^{0\centerdot
}_{0,p}{ C}_{0,p}^\otimes ) ={ L}^{wr}_{0,p}{\it \Phi}_{0,p}^{wr}$.

The matrices of the kind ${ L}^{wr}_{0,p}$ 
form a group because there are $L^{{wr}}_{0,p}L^{{wr}}_{p,k}=T_{0,p}^0C_{0,p}^\otimes T_{p,k}^pC_{p,k}^\otimes=$

$T_{0,p}^0T_{p,k}^0C_{0,p}^{\otimes} C_{p,k}^\otimes =T^0_{0,k}C_{0,k}^\otimes=L^{{wr}}_{0,k}$ for a subindex $p$ and $%
L_{0,k}^{{{wr}},-1}=(T^0_{0,k}C_{0,k}^{\otimes})^{-1}=C_{0,k}^{\otimes,T}(T^0_{0,k})^{-1}
$

$=C_{k,0}^\otimes T^0_{k,0}C_{0,k}^{\otimes,T}C_{k,0}^\otimes =T^k_{k,0}C_{k,0}^\otimes=L^{{wr}}_{k,0}$.

The similar statement is true for twists $\pi ^{tw}$: $\pi_0 ^{tw, 0}= {L}^{tw}_{0,p}\pi_p^{tw ,p}$ where we have the matrix  
${ L}^{tw}_{0,p}=\begin{bmatrix} O &\ I \\ I 
&\ O \end{bmatrix}{ L}^{wr}_{0,p}\begin{bmatrix} O &\ I \\ I 
&\ O \end{bmatrix}$ belongs to the multiplicative group ${\mathcal L}^{tw}({\mathcal R}, 6)$ such that $
{ L}^{{tw}\centerdot }_{0,p}={ L}^{tw}_{0,p}{\it \Phi}_{0,p}^{tw}$, 
$ {\it \Phi}_{0,p}^{tw}=- {\it\Phi}_{0,p}^{wr, T}
$. 

Note that in contrast to the groups of motions in the $3$--dimensional space the groups ${\mathcal L}^{wr}({\mathcal R}, 6)$ and ${\mathcal L}^{tw}({\mathcal R}, 6)$ are multiplicative. \medskip 

\noindent{\bf Newton--Euler equation}. Let the frame ${\cal E}_{p}$ be attached to a body $A_p$. 

\begin{Lemma} {\rm \cite{Konoplev1999}}  {
There is the following relation} 
\[
{l}_p^{v_{x},p}={\it \Theta }_p^xV_{0,p}^p,\quad {\it \Theta }_p^x= 
\begin{bmatrix}
I & -r_p^{p\times} \\ 
 r_p^{p\times} & -(r_p^{p\times} )^2
\end{bmatrix}
\]
\end{Lemma}
\proof
The statement is true as 
\[
{l}_p^{v_{x},p} =
\begin{bmatrix}
I \\ 
 r_p^{p\times}
\end{bmatrix} v_{x}^p=
\begin{bmatrix}
I \\ 
 r_p^{p\times} 
\end{bmatrix} \left( v_{0,p}^p+ \omega_{0,p}^{p\times} r_p^p\right) = 
\begin{bmatrix}
I & -r_p^{p\times} \\ 
 r_p^{p\times} & -(r_p^{p\times} )^2
\end{bmatrix}  
\begin{pmatrix}
v_{0,p}^p \\ 
\omega_{0,p}^p
\end{pmatrix}
\]
where the relation $\omega_{0,p}^{p\times} r_p^p=-r_p^{p\times} \omega_{0,p}^p$ is used.

Due to the lemma  we have $
\left({l}_0 ^{ v_{x},0}\right)^{\centerdot  }=\left({ L}^{wr}_{0,p}{l}_p^{ v_x,p}\right)^{\centerdot  }=
 { L}^{wr}_{0,p}({ \it \Theta }_p^x V_{0,p}^{p\centerdot  }+{ \it \Phi}_{0,p}^{wr}{ \it \Theta }_p^x V_{0,p}^p)
$ or ${ L}^{wr}_{p,0}
\left({l}_0 ^{ v_{x},0}\right)^{\centerdot  }=
 ({ \it \Theta }_p^x V_{0,p}^{p\centerdot  }+{ \it \Phi}_{0,p}^{wr}{ \it \Theta }_p^x V_{0,p}^p)
$. That is why from (\ref{ 13a})--(\ref{ 13b}) follows that
	\[\int \hspace{-.05cm} \chi _{_{A_p}}
\rho_x ({ \it \Theta }_p^x V_{0,p}^{p\centerdot  }+{ \it \Phi}_{0,p}^{wr}{ \it \Theta }_p^x V_{0,p}^p)\mu_{_{LS}} (dx)=\int \hspace{-.05cm} \chi _{_{A_p}}{ L}^{wr}_{p,0}[l_0^{\phi(x,x^e),0}+l_0^{c(x,x^e),0}]\mu_{_{LS}} (dx)
\]
where (and henceforth) all integrals are taken w.r.t. Lebesgue--Stieltjes  measure $\mu_{_{_{LS}}}$; the set $A_p$ is immobile in the frame ${\cal E}_p$.

According the rigid body definition the constraints are considered as ideal and thus \cite{Vilke} 
	\[\int \hspace{-.05cm} \chi _{_{A_p}}{ L}^{wr}_{p,0}l_0^{c(x,x^e),0}\mu_{_{LS}} (dx)=0
\]
\begin{Theorem} {\rm \cite{Konoplev1999}} 
The motion of $A_p$ (w.r.t. ${\cal E}_0$ in the frame ${\cal E}_p$) is
described by the ({\em Newton--Euler}) equation (see also \cite{Berthelot})
\begin{equation}{ \it \Theta }_p V_{0,p}^{p\centerdot  }+{ \it \Phi}^{wr}_{0,p}{ \it \Theta }_p V_{0,p}^p ={\mathcal F}_{0}^p
\label{ 6g}
\end{equation}
where ${ \it \Theta }_p=\hspace{-0.1cm} \int \hspace{-.05cm} \chi _{_{A_p}} {\it \Theta }_p^x \rho _x \mu_{_{LS}}(dx)$, ${\mathcal F}^p_{0}=\int \hspace{-.05cm} \chi _{_{A_p}}l_0^{\phi(x,x^e),p}\mu_{_{LS}}(dx)$ is the wrench calculated in ${\cal E}_p$ and generated by the main vector and the total moment acting on the body $A_p$.
\end{Theorem}
\noindent{\bf Systems of consecutively connected bodies  {\rm \cite{Konoplev2010}}}. Let us consider a system of $k+1$ consecutively connected bodies $A_p$, $p=\overline{0,k}$ (the body $A_0$ is immobile). Its motion is depicted by the following Newton--Euler equation
\begin{equation}
	A V_a^{ \centerdot}+B V_a= F_a
	\label{ne}
\end{equation}
where $A$ and $B$ are known matrices, $V_a={\rm col}\{V_{0,{p}}^{p}\}$, $F_a={\rm col}\{F_{0}^{p}\}$, $p=\overline{1,k}$.

 Newton--Euler equation (\ref{ne}) is considered w.r.t.  `absolute' quasi--velocities $V_{0,{p}}^{p}$ of the bodies (calculated in ${\mathcal E}_p$ w.r.t. the main frame ${\mathcal E}_0$). But in practice there are only the `relative' quasi--velocities $V_{p-1,p}$ of the frame ${\mathcal E}%
_p$ w.r.t. ${\mathcal E}_{p-1}$. 
 Thus we must connect the `absolute' quasi--velocities with `relative' ones.\medskip

\begin{Lemma} {\em \cite{Konoplev1999}} 
For a system of consecutively connected bodies there is the following composition rule
	\[	\begin{pmatrix}\vec{v}_{0,p}\\
	\vec{\omega}_{0,p}\end{pmatrix}=
	\begin{pmatrix}\vec{v}_{0,p-1}\\
	\vec{\omega}_{0,p-1}\end{pmatrix}+
	\begin{pmatrix}\vec{v}_{p-1,p}\\
	\vec{\omega}_{p-1,p}\end{pmatrix}\]
\end{Lemma}
\proof 
As the bodies are connected consecutively there is the relation 
${ C}_{0,p}={ C}_{0,p-1}{ C}_{p-1,p}
$ is true. With differentiating it we have 
 $\omega _{0,p}^{p}=\omega _{0,p-1}^{p}+\omega
_{p-1,p}^{p}=C_{p,p-1}\omega _{0,p-1}^{{p}-1}+\omega _{p-1,p}^{p}
$. 
Besides define the vectors $\vec{d}_{0,p-1}=\overrightarrow{(O_0,O_{{p}-1})}$ and $\vec{d}_{{p-1,p}}=\overrightarrow{(O_{{p}-1},O_{p})}$, then $
d_{0,p}^0=d_{0,p-1}^0+d_{p-1,p}^0$, $v_{0,p}^0=v_{0,p-1}^0+d_{p-1,p}^{0%
\centerdot }$, $d_{p-1,p}^0=C_{0,p-1}d_{p-1,p}^{{p}-1}$, $d_{p-1,p}^{0\centerdot
}=v_{p-1,p}^0+C_{0,p-1}^{\centerdot }d_{p-1,p}^{{p}-1}=C_{0,p-1}\omega _{0,p-1}^{{p}-1\times
}d_{p-1,p}^{{p}-1}+v_{p-1,p}^0=v_{p-1,p}^0+C_{0,p-1}\omega _{0,p-1}^{{p}-1\times
}C_{0,p-1}d_{p-1,p}^{{p}-1}=v_{p-1,p}^0-d_{p-1,p}^{0\times }\omega _{0,p-1}^0=v_{p-1,p}^0+d_{p,p-1}^{0\times }\omega _{0,p-1}^0$. Hence $v_{0,p}^0=v_{0,p-1}^0+v_{p-1,p}^0+d_{p,p-1}^{0\times }\omega
_{p-1,p}^0$, $v_{0,p}^{p}=C_{p,p-1}v_{0,p-1}^{{p}-1}+C_{p,p-1}d_{p,p-1}^{{p}-1\times
}\omega _{0,p-1}^{{p}-1}+v_{p-1,p}^{p}$,  $
V_{0,p}^{p}=
\begin{bmatrix}
C_{p,p-1} & \hspace{-.2cm}O \\ 
O &\hspace{-.2cm} C_{p,p-1}
\end{bmatrix}\hspace{-.2cm}
\begin{bmatrix}
I &d_{p,p-1}^{p-1\times } \\ 
O& I
\end{bmatrix}V_{0,p-1}^{p-1}+V_{p-1,p}^{p}=L_{p,p-1}^{tw}V_{0,p-1}^{p-1} +V_{p-1,p}^{p}$. 

Hence we have\vspace{-7pt}
\begin{equation}\quad 
V_{0,p}^{p}=\sum_{k=1}^{k=p}L_{p,k}^{tw}V_{k-1,k}^{k}, \quad V_{k-1,k}^{k}= 
\begin{pmatrix}
v_{k-1,k}^k \\ 
\omega _{k-1,k}^k
\end{pmatrix}
\label{ 16}
\end{equation}
where ${ L}_{p,k}^{tw}=\begin{bmatrix}
C_{p,k} & O \\ 
O & C_{p,k}
\end{bmatrix} 
\begin{bmatrix}
I &\hspace{ .015cm}d_{p,k}^{k\times } \\ 
O& \hspace{ .015cm}I
\end{bmatrix}$, 
$L^{tw}_{k,k}=I$.

From (\ref{ 16}) follows the {\em equation of kinematics}\index{Equation!of kinematics} 
\begin{equation}\quad 
{ V}_a={ L} { V}_{r}
\label{ 17}
\end{equation}
where ${V}_a\hspace{-.1cm}=\hspace{-.1cm}{\rm  col}%
\{{\it V}_{0,1}^1,\ldots ,\hspace{-.1cm} {\it V}_{0,\it k}^k,\ldots ,\hspace{-.1cm}{\it V}_{ 0,{\it n}}^n\}$, ${ V}_r\hspace{-.1cm}=\hspace{-.1cm}{\rm  col}%
\{{\it V}_{0,1}^1,\ldots ,\hspace{-.1cm}{\it V}_{{\it k}-1,\it k}^{k-1},\ldots ,\hspace{-.1cm}{\it V}_{ {\it n}-1,{\it n}}^n\}$, ${ L}$ is the triangular matrix with blocks ${ L}_{p,k}^{tw}$ being functions of `relative' frame rotations and translations.

Thus we have 
\begin{equation}
	A L V_r^{ \centerdot}+(A L^{ \centerdot} +B)V_r= F_a \label{NewtonE}
\end{equation}
where $L^{\centerdot}$ is analytically calculated due to relation (\ref{ 4}).

It is easy to see that the matrices of relation (\ref{NewtonE}) depend on rotation matrices (and linear and angular quasi--velocities, too) that is why equation (\ref{NewtonE}) must be considered along with the {\em Euler kinematical relation}
\begin{equation}
	C_{p-1,p}^{\centerdot  } =
	C_{p-1,p}\omega_{p-1,p}^{p\times}\label{Ek}
\end{equation}\vspace{7pt}

\begin{figure}[ht]
\begin{center}
\hspace{-.06cm}\includegraphics[width=9truecm, bb=0 0 1416 759]{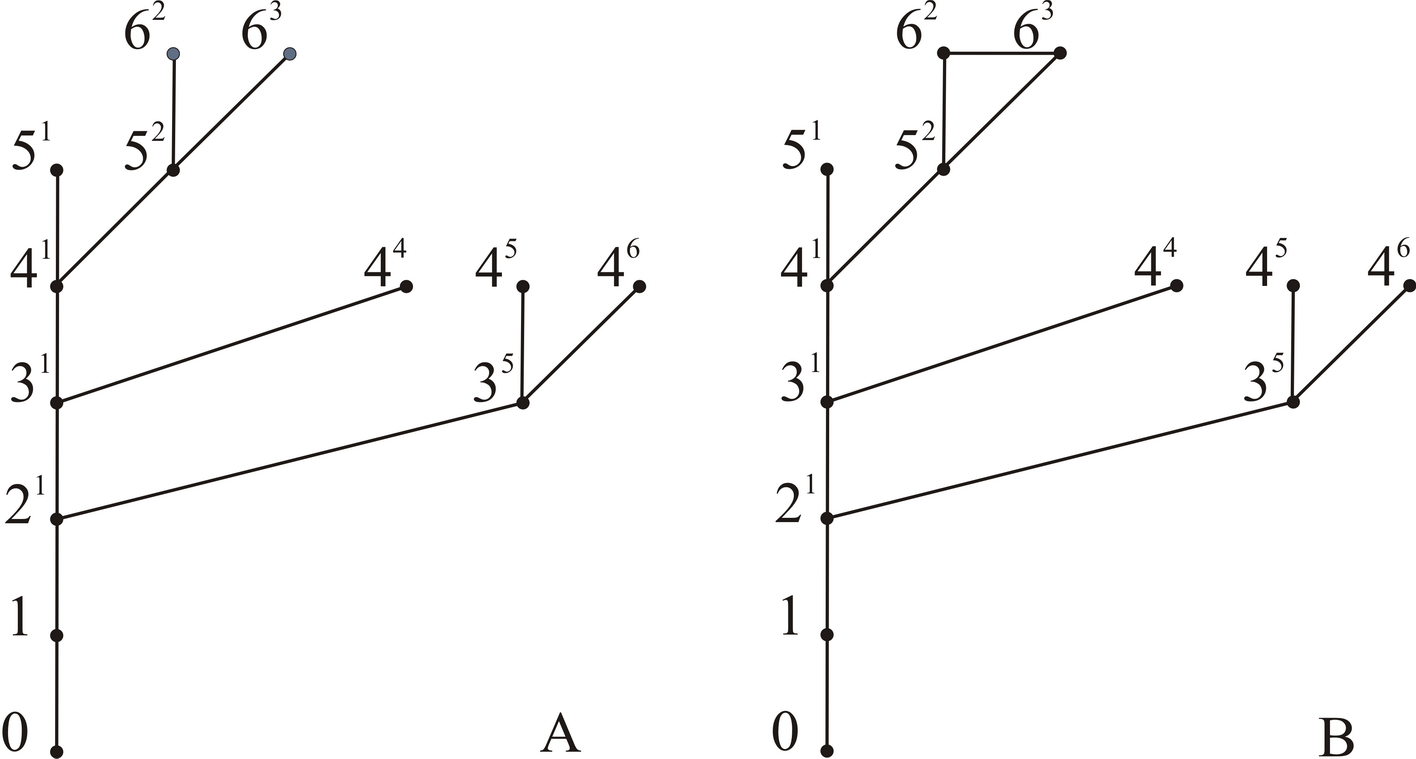}
\end{center}
\begin{center}%
{Fig. 1. Multibody system graphs}\end{center}
 \vspace{-.5cm} \end{figure}
  
  \noindent{\bf Multibody systems with tree--like structure}.
Consider a multibody system with tree--like structure given by 
 the graph in Fig. 1A.
  Let vertices $j^i$ represent the system bodies or the origins of the attached Cartesian frames ${{\mathcal E}}_j^i$ where the index $i$ numbers the tree--tops, the index $j$ numbers the bodies from the base to the corresponding tree--tops. Introduce $V_{k,j}^{m,i}$ as quasi--velocities characterizing rotation and translation of the frames ${{\mathcal E}}_j^i$ w.r.t. ${{\mathcal E}}_k^m$. Then we have the sets 
 $\{V_{0,\hspace{-.01cm}1}^{0,\hspace{-.01cm}1}, V_{1,\hspace{-.01cm}2}^{1,\hspace{-.01cm}1}, V_{2,\hspace{-.01cm}3}^{1,\hspace{-.01cm}1}, V_{3,\hspace{-.01cm}4}^{1,\hspace{-.01cm}1}, V_{4,\hspace{-.01cm}5}^{1,\hspace{-.01cm}1}\}, 
\{V_{0,\hspace{-.01cm}1}^{0,\hspace{-.01cm}1},\ V_{1,\hspace{-.01cm}2}^{1,\hspace{-.01cm}1}, V_{2,\hspace{-.01cm}3}^{1,\hspace{-.01cm}1}, V_{3,\hspace{-.01cm}4}^{1,\hspace{-.01cm}1},
V_{4,\hspace{-.01cm}5}^{1,\hspace{-.01cm}2}, V_{5,\hspace{-.01cm}6}^{2,\hspace{-.01cm}2}\},\
\{V_{0,\hspace{-.01cm}1}^{0,\hspace{-.01cm}1},
V_{1,\hspace{-.01cm}2}^{1,\hspace{-.01cm}1},
V_{2,\hspace{-.01cm}3}^{1,\hspace{-.01cm}1},$
 
\noindent 
$ V_{3,\hspace{-.01cm}4}^{1,\hspace{-.01cm}1},
V_{4,\hspace{-.01cm}5}^{1,\hspace{-.01cm}2}, V_{5,\hspace{-.01cm}6}^{2,\hspace{-.01cm}3}\},\
\{V_{0,\hspace{-.01cm}1}^{0,\hspace{-.01cm}1} V_{1,\hspace{-.01cm}2}^{1,\hspace{-.01cm}1}, V_{2,\hspace{-.01cm}3}^{1,\hspace{-.01cm}1}, V_{3,\hspace{-.01cm}4}^{1,\hspace{-.01cm}4}\},\
\{V_{0,\hspace{-.01cm}1}^{0,\hspace{-.01cm}1} V_{1,\hspace{-.01cm}2}^{1,\hspace{-.01cm}1} V_{2,\hspace{-.01cm}3}^{1,\hspace{-.01cm}5} V_{3,\hspace{-.01cm}4}^{5,\hspace{-.01cm}5}\},\
\{V_{0,\hspace{-.01cm}1}^{0,\hspace{-.01cm}1}, V_{1,\hspace{-.01cm}2}^{1,\hspace{-.01cm}1}, V_{2,\hspace{-.01cm}3}^{1,\hspace{-.01cm}5},
V_{3,\hspace{-.01cm}4}^{5,\hspace{-.01cm}6}\}$ 
and $\{V_{0,1}^{0,1}$,

 \noindent 
$
V_{0,2}^{1,1}\hspace{-.01cm},\hspace{-.01cm} V_{0,3}^{1,1}\hspace{-.01cm},\hspace{-.01cm} V_{0,4}^{1,1}, V_{0,5}^{1,1}\},
\{V_{0,1}^{0,1}\hspace{-.01cm},\hspace{-.01cm} V_{0,2}^{1,1}\hspace{-.01cm},\hspace{-.01cm}V_{0,3}^{1,1}\hspace{-.01cm},\hspace{-.01cm} V_{0,4}^{1,1}\hspace{-.01cm},\hspace{-.01cm}V_{0,\hspace{-.01cm}5}^{1,\hspace{-.01cm}2}\hspace{-.01cm},\hspace{-.01cm}
V_{0,\hspace{-.01cm}6}^{2,\hspace{-.01cm}2}\},
\{V_{0,\hspace{-.01cm}1}^{0,\hspace{-.01cm}1}\hspace{-.01cm},\hspace{-.01cm}V_{0,\hspace{-.01cm}2}^{1,\hspace{-.01cm}1}\hspace{-.01cm},\hspace{-.01cm} V_{0,\hspace{-.01cm}3}^{1,\hspace{-.01cm}1}\hspace{-.01cm},\hspace{-.01cm} V_{0,\hspace{-.01cm}4}^{1,\hspace{-.01cm}1}\hspace{-.01cm},\hspace{-.01cm} V_{0,\hspace{-.01cm}5}^{1,\hspace{-.01cm}2}\hspace{-.02cm},\hspace{-.01cm}V_{0,\hspace{-.01cm}6}^{2,\hspace{-.01cm}3}\},\{V_{0,\hspace{-.01cm}1}^{0,\hspace{-.01cm}1}\hspace{-.01cm},\hspace{-.01cm}$

 \noindent $
V_{0,\hspace{-.01cm}2}^{1,\hspace{-.01cm}1}\hspace{-.01cm},\hspace{-.01cm} V_{0,\hspace{-.01cm}3}^{1,\hspace{-.01cm}1}\hspace{-.02cm},\hspace{-.01cm} V_{0,\hspace{-.01cm}4}^{1,\hspace{-.01cm}4}\},
\{V_{0,\hspace{-.01cm}1}^{0,\hspace{-.01cm}1}\hspace{-.01cm},\hspace{-.01cm} V_{0,\hspace{-.01cm}2}^{1,\hspace{-.01cm}1},V_{0,\hspace{-.01cm}3}^{1,\hspace{-.01cm}5}\hspace{-.01cm},\hspace{-.01cm}
V_{0,4}^{5,5}\}\hspace{-.01cm},\hspace{-.01cm}
\{V_{0,1}^{0,1}\hspace{-.02cm},\hspace{-.01cm}V_{0,2}^{1,1}\hspace{-.01cm},\hspace{-.01cm} V_{0,3}^{1,5}\hspace{-.01cm},\hspace{-.01cm} V_{0,4}^{5,6}\}$ 
with the same subscripts as in the case of consecutively connected bodies for the relative and absolute quasi--velocities. 

 This case is considered above that is why we arrive at relation 
(\ref{ 17})  with the known matrix ${L}$ and 
 \begin{eqnarray*}
{ V}_a &=& {\rm  col}\{
V_{0,1}^{0,1}, V_{0,2}^{1,1}, V_{0,3}^{1,1}, V_{0,4}^{1,1}, V_{0,5}^{1,1}
, V_{0,5}^{1,2}, V_{0,6}^{2,2}
,V_{0,6}^{2,3},V_{0,4}^{1,4},
V_{0,3}^{1,5}, V_{0,4}^{5,5}
,V_{0,4}^{5,6}\}\\
{ V}_r  &=& 
{\rm  col}\{
V_{0,1}^{0,1}, V_{1,2}^{1,1}, V_{2,3}^{1,1}, V_{3,4}^{1,1}, V_{4,5}^{1,1}, 
V_{4,5}^{1,2}, V_{5,6}^{2,2},
V_{5,6}^{2,3},V_{3,4}^{1,4},
V_{2,3}^{1,5}, V_{3,4}^{5,5},
V_{3,4}^{5,6}\}
\end{eqnarray*}

\begin{Remark}The results obtained can be immediately applied to systems with loops, {\em e.g.}, if in the system under consideration (see Fig. 1B) the vertex $6^2$ is connected with $6^3$ by the edge ($6^2,6^3$). In this case relation (\ref{ 6g}) is the same, but in the case where constraints are considered there are the following additional constraints $\overrightarrow{(5^2,6^2)}+\overrightarrow{(6^2,6^3)}+\overrightarrow{(6^2,5^2)}=0$ and $C^{2,2}_{5,6}C^{2,3}_{6,6}C^{2,2}_{6,5}=I$.
\end{Remark}

{\bf Parameterization of rotation matrices}. 
The order of system (\ref{NewtonE})--(\ref{Ek}) may be reduced. To this end one uses different parameterizations of rotation matrices. \medskip

\underline{\em Euler angles}. Let $C_{p-1,{p}}=C_1C_2C_3$ where \begin{equation}
C_1 =
\begin{bmatrix}\hspace{.1cm}
1\  & 0 & 0  \\ \hspace{.1cm}
0\  & \cos \varphi  & -\sin
\varphi  \\ \hspace{.1cm} 
0\ &  \sin \varphi  & %
\cos \varphi 
\end{bmatrix} ,\, C_2 =
\begin{bmatrix}
\cos \vartheta  &\ 0 &  \sin \vartheta  \\ 
0 &\ 1 & 0
\\ 
-\sin \vartheta  &\ 0 & %
\cos \varphi 
\end{bmatrix} ,\, C_3 =
\begin{bmatrix}\hspace{.1cm}
\cos \psi  & -\sin \psi \hskip %
-.04in &\ 0 \hspace{.1cm}  \\  \hspace{.1cm} 
 \sin \psi  & \cos \psi 
&\ 0 \hspace{.1cm}  \\  \hspace{.1cm} 
0 & 0 &\ 1\hspace{.1cm}
\end{bmatrix}
\label{ 26}
\end{equation}
are so called the simplest rotation\index{Rotation!simplest} matrices;  $\varphi $, $\vartheta $, and $\psi $ are  
Euler angles \cite{Korn}.

Introduce the triple $ \lambda_{p-1,{p}}={\rm col}\{\varphi , \vartheta , \psi\}$ as a parameter. Then there is the matrix $D_{p-1,{p}}$ such that \cite{Konoplev1996}
\begin{equation}\omega _{p-1,{p}}^{p}=D_{p-1,{p}}\lambda_{{p-1,{p}}}^{\centerdot}
\label{ 7g}\end{equation}

Hence equation (\ref{NewtonE}) must be considered along with the following  relation
\begin{equation}\lambda_{{p-1,{p}}}^{\centerdot}=D_{p-1,{p}}^{-1}\omega _{p-1,{p}}^{p}
\label{ 7ghh}\end{equation}
and $C_{p-1,{p}}=C_{p-1,{p}}(\lambda_{{p-1,{p}}})$ if the matrix $D_{p-1,{p}}^{-1}$ exists.\medskip

\underline{\em   Fedorov vector--parameter}. To parameterize rotation matrices we may introduce Fedorov vector--parameter \cite{Fedorov}.
\begin{Definition}{\rm \cite{Fedorov}} 
The number triple $f\in {\bf R}_3$ is called {\em Fedorov vector--parameter} of a rotation matrix $C$, if it is answered with the following matrix (see (\ref{ 2}))
\[
 f^{\times}= 
(C-I)(C+I)^{-1}
\]
\end{Definition}

The inverse map of Cayley restores the rotation matrix 
\[
 C{=} (I+f^{\times})(I-f^{\times})^{-1}
\]
It is easy to be verified (for example, by means of Maple$^ {\copyright} $) that 
 the following relations are true
\begin{equation}
	f^{\times } =\frac{C-C^T}{1+{\rm  tr}\hspace{0.05cm}\mathit{C}}
\label{map1}\vspace{7pt}
\end{equation}
\begin{equation}
	C =\frac{(1-\Vert f \Vert ^2)I+2f f^T+2f^{\times }}{1+\Vert f \Vert ^2}
\label{map2}\vspace{-.3in}
\end{equation}\vspace{11pt}

Let the rotation matrices $C_{0, p}$ and $C_{p, k}$ have Fedorov vector--parameters $f_{0, p}$ and $f_{p, k}$. It is known that they are eigenvectors of these matrices, {\em i.e.},
\[C_{0, p}f_{0, p}=f_{0, p}, \quad 
C_{p, k}f_{p, k}=f_{p, k} \in {\bf R}_3
\]
As the space ${\bf R}_3$ has 3 bases ${\bf e}^0$, ${\bf e}^{p}$ and ${\bf e}^k$ we may write  $f_{0, p}=f_{0, p}^0=f_{0, p}^{p}$, $ f_{p, k}=f_{p, k}^{p}=f_{p, k}^k$ 
and define the following vector 
\[
\vec r_{p, k}= \sum_i f_i \vec e_i^{\hspace{.1cm } p}=\sum_i f_i \vec e_i^{\hspace{.1cm } k}=\sum_i f_i^0 \vec e_i^{\hspace{.1cm } 0}
\]
where 
${\rm col}\{f_1, f_2, f_3\}=f_{p, k}^{p}$ and ${\rm col}\{f_1^0, f_2^0, f_3^0\}=C_{0, p}f_{p, k}^{p}$.
\begin{Definition}
The vector 
$\vec r_{p, k}$ is called vector of {\em Rodrigues} (the half of the vector of finite rotation -- see \cite{Lurie}).
\end{Definition}
 
The definition is  motivated by the fact that so defined vector $\vec r_{p, k}$ is collinear with the instantaneous angular
velocity $\vec{\omega}_{p,k}$, and thus it defines the rotation axis.

\vspace{.3cm}

\begin{figure}[ht]
\begin{center}
\hspace{-.06cm}\includegraphics[width=8truecm, bb=0 0 1417 683]{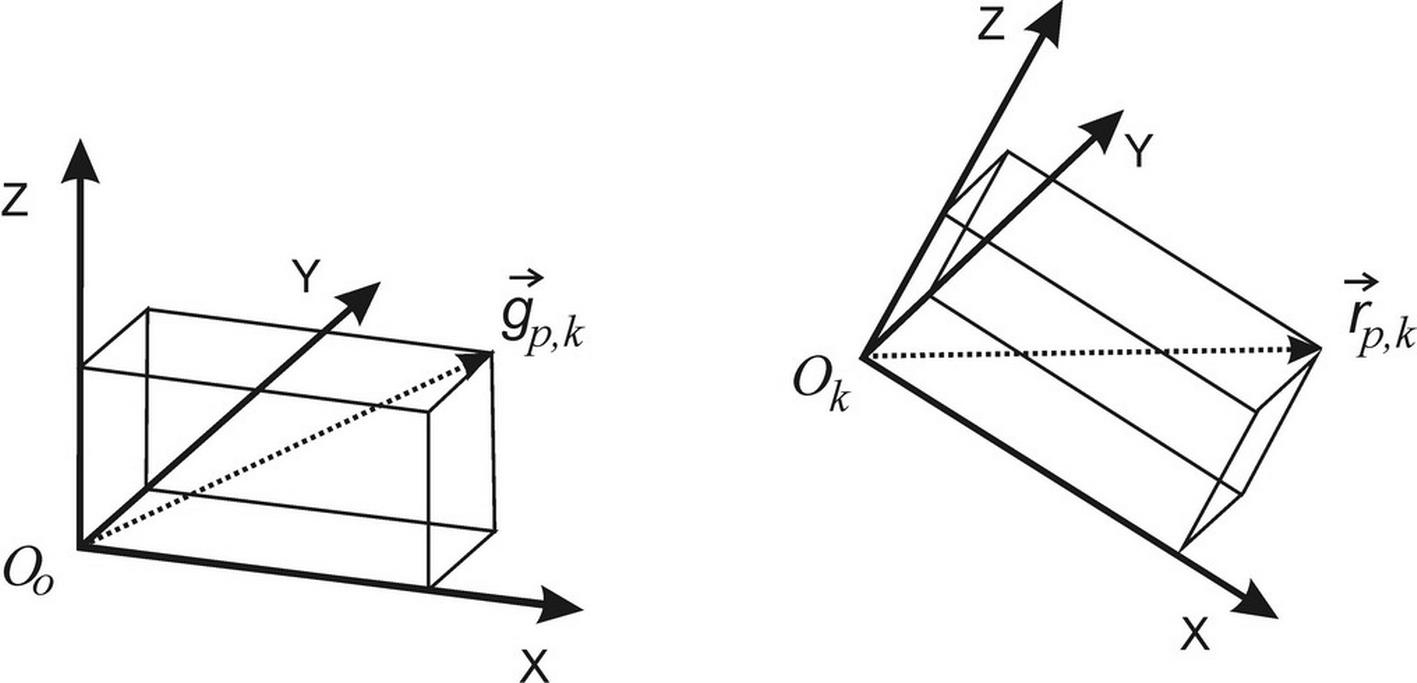}
\end{center}
\begin{center}%
{Fig. 2. Gibbs and Rodrigues vectors.}\end{center}
 \vspace{-.5cm} \end{figure}  
\begin{Remark}In \cite{Fedorov} it is explicitly pointed out that vector--parameters are {\em Gibbs} vectors which can be defined in the form $\vec g_{p, k}=\sum_i f_i \vec e_i^{\hspace{.1cm } 0}$ (in mechanics free vectors and their coordinate columns in the canonical basis ${\bf e}^0$ are known with the same name {\em vector} as elements of the vector spaces ${\bf V}_3$ and ${\bf R}_3$, and there is no other basis except the canonical one in \cite{Fedorov}). In general the vector $\vec g_{p, k}$ does not coincide with  $\vec r_{p, k}$  (see Fig. 2), as  the bases ${\bf e}^0$, ${\bf e}^{p}$ and ${\bf e}^k$ are different.
\end{Remark}
 
There is the following relation \cite{Lurie} 
\[ 
\vec{\omega}_{p-1,{p}}=\frac 2{1+\Vert \vec{r}_{p-1,{p}}\Vert ^2}( \vec{r}^{\hspace{ .08cm}\centerdot }_{p-1,{p}}+
\vec{r}_{p-1,{p}}\times \vec{r}^{\hspace{ .08cm}\centerdot }_{p-1,{p}})
\]
 As the vectors $\vec{\omega}_{p-1,{p}}$ and $\vec{r}_{p-1,{p}}$ are collinear, we have relation (\ref{ 7g})  
where  $\lambda_{{p-1,{p}}}={f}_{p-1,{p}}$ and  $D_{p-1,{p}}=\frac 2{1+\Vert {f}_{p-1,{p}}\Vert ^2}(I+
{f}_{p-1,{p}}^{\times})$.

Thus equation (\ref{NewtonE}) must be considered along with the following  relation
\begin{equation}
f_{p-1,p}^{\centerdot}=\frac 12\, (1+\Vert {f}_{p-1,p}\Vert^2)(I-f_{p-1,{p}}^{\times})^2(I+f^{\times}_{p-1,{p}})\omega _{p-1,{p}}^{p}
\label{ 7ghha}\end{equation}
and $C_{p-1,{p}}=C_{p-1,{p}}(f_{{p-1,{p}}})$.
\medskip 

\underline{\em   Euler--Rodrigues  parameters}. To parameterize rotation matrices we may use quaternions. 
\begin{Definition}The set $\Lambda=\{{\lambda_0}\in {\bf R}, \vec \lambda\in {\bf V}_3\}$ is called {\em quaternion}.
\end{Definition} Quaternions generate the algebra with the quaternion product 
\[\Lambda \circ M=\{{\lambda_0}{\mu_0}-\langle \vec \lambda,\vec {\mu} \rangle , {\lambda_0}\vec \mu+{\mu_0}\vec \lambda+ \vec \lambda\times \vec \mu\}
\]
where $M=\{{\mu_0}, \vec \mu\}$, $\langle \vec \lambda,\vec {\mu} \rangle$ is the inner product. 


Any vector $\vec \lambda$ can be imaged as a quaternion $\Lambda$ with the zero scalar part. That is why we  may define the quaternion product of two vectors $\vec \lambda$ and $\vec {\mu}$ as follows 
\[\vec \lambda\circ \vec {\mu}=\{-\langle \vec \lambda,\vec {\mu} \rangle,  \vec \lambda\times \vec \mu\}
\]

There exists the unit quaternion ${\Lambda}_{p-1,{p}}=\{{\lambda_0}, \vec \lambda_{p-1,{p}}\}$ (with $\Vert\Lambda_{p-1,{p}}\Vert = 1$) such that \cite{Yakovenko}
\begin{equation}
{\Lambda}_{p-1,{p}}\circ\vec \omega_{p-1,{p}} \circ {\Lambda}_{p-1,{p}}= \vec \omega_{p-1,{p}},\quad \vec \omega_{p-1,{p}} =-2{\Lambda}_{p-1,{p}}\circ \widetilde{\Lambda}_{p-1,{p}}^{\centerdot},\quad 
{\Lambda}_{p-1,{p}}^{\centerdot}=\frac 12\, \vec \omega_{p-1,{p}} \circ {\Lambda}_{p-1,{p}}
\label{13cc}
\end{equation}
where $\widetilde{\Lambda}_{p-1,{p}}=\{{\lambda_0}, -\vec \lambda_{p-1,{p}}\}$ is  {\em conjugation} of $\Lambda_{p-1,{p}}$.

Let us denote  ${\rm col}\{\omega _1, \omega_2, \omega_ 3\}\stackrel{\mathrm{def}}{=}\omega _{p-1,{p}}^{p}$, ${\rm col}\{\lambda  _1, \lambda _2, \lambda _ 3\}\stackrel{\mathrm{def}}{=}\lambda _{p-1,{p}}^{p}$ and ${\rm col}\{\lambda  _0, \lambda _{p-1,{p}}^{p}\}\stackrel{\mathrm{def}}{=}\Lambda _{p-1,{p}}^{p}$
 then 
the orthogonal matrix $C_{p-1,{p}}$ corresponding to a rotation by the unit quaternion $\Lambda_{p-1,{p}}$  is given in the following form \cite{Yakovenko}
\begin{equation}C_{p-1,{p}}(\Lambda_{p-1,{p}}^{p})=
	\begin{bmatrix}
\lambda_0^2+\lambda_1^2-\lambda_2^2-\lambda_3^2&2\lambda_1\lambda_2-2\lambda_0\lambda_3        &2\lambda_1\lambda_3+2\lambda_0\lambda_2        \\
	2\lambda_1\lambda_2+2\lambda_0\lambda_3 &\lambda_0^2-\lambda_1^2+\lambda_2^2-\lambda_3^2&2\lambda_2\lambda_3-2\lambda_0\lambda_1        \\
	2\lambda_1\lambda_3-2\lambda_0\lambda_2        &2\lambda_2\lambda_3+2\lambda_0\lambda_1        &\lambda_0^2-\lambda_1^2-\lambda_2^2+\lambda_3^2\\
	\end{bmatrix}
	\label{RHm}\end{equation}
The quadruple $\Lambda_{p-1,{p}}^p$ is known as that of Euler-Rodrigues  parameters.

From (\ref{13cc}) follows \cite{Yakovenko}
\begin{equation}
	\Lambda_{p-1,{p}}^{p\centerdot} =\frac 12\, 
	\begin{bmatrix}
	0 & -\omega _1 & -\omega _2 & -\omega _3 \\ 
	\omega _1 & 0 & \omega _3 & -\omega _2 \\ 
	\omega _2 & -\omega _3 & 0 & \omega _1 \\ 
	\omega _3 & \omega _2 & -\omega _1 & 0
	\end{bmatrix} \Lambda_{p-1,{p}}^p
	\label{RH}
\end{equation}
 Hence equation (\ref{NewtonE}) must be considered along with  relations (\ref{13cc})--(\ref{RH}).

From (\ref{RHm}) follows also that there is the matrix $D_{p-1,{p}}=D_{p-1,{p}}(\Lambda_{p-1,{p}}^p)$ such that relation $\omega _{p-1,{p}}^{p}=D_{p-1,{p}}\Lambda_{{p-1,{p}}}^{p\centerdot}$ is true.\medskip

{\bf Lagrange equation of II kind}. Let $\lambda _{p-1,{p}}$ be a triple of Euler angles or Fedorov vector--parameter.

\begin{Definition} \begin{enumerate}
\item  The vectors $q_{p-1,{p}}={\rm  col}\{{\it d}_{p-1,{p}}^{\it p},\lambda _{p-1,{p}}\}$ and ${\it q}_{p-1,{p}}^{\centerdot }={\rm  col}\{{\it d}_{p-1,{p}}^{{\it p}\centerdot },\lambda^{\centerdot }_{p-1,{p}}\} 
$ are called {\em canonical generalized coordinates} and {\em velocities} of the frame 
${\cal E}_p$ in the motion w.r.t. the frame ${\cal E}_{p-1}$;\vspace{-7pt}
\item  the relation \vspace{-21pt}\end{enumerate}
\begin{equation}
V_{p-1,{p}}^p=M_{p-1,{p}} q_{p-1,{p}}^{\centerdot },\quad M_{p-1,{p}}={\rm  diag}\{ I,D_{p-1,{p}}\}
\label{ 1011}\vspace{-3pt}
\end{equation}
\hspace{1cm} is called {\em equation of kinematics}\index{Equation!of kinematics} of 
${\cal E}_p$--frame w.r.t. ${\cal E}_{p-1}$.\vspace{-7pt}
\end{Definition}

From relations (\ref{NewtonE}) and (\ref{ 7g}) follows the {\em Lagrange equation}\index{Equation!of Lagrange}  of II kind 
\begin{equation} 
{\mathcal A}(q)q^{\centerdot \centerdot }+{\mathcal B}(q,q^{\centerdot })
q^{\centerdot }={\mathcal F}
\label{ 23}
\end{equation}
where ${\mathcal A}(q)=L^TM^TA LM$, ${\mathcal B}(q,q^{\centerdot })=
L^TM^T[A L M^{\centerdot} +(A L^{ \centerdot} +B)M]$, ${\mathcal F}=L^TM^TF_a$, $M={\rm diag}\{M_{p-1,{p}}\}$, $q={\rm col}\{q_{p-1,{p}}\}$.


In the many cases there are constraints on motion of multibody systems, and the matrix $N$ exists such that the matrix $N^TN$ is non--degenerate  and we may introduce the generalized coordinate $q_c=Nq \in {\bf R}_{m}$ where the natural number $m$ is not more $6k$ \cite{Vel}. 
 Then from relation (\ref{ 23}) follows 
\[{\cal A}_c q_c^{\centerdot \centerdot}+{\cal B}_cq_c^{\centerdot}= {\cal F}_c
\]
where ${\cal A}_c$, ${\cal B}_c$ and ${\cal F}_c$ are known matrices and column.

As to the quadruple $\Lambda_{p-1,{p}}^p$, we may replace $\lambda_{p-1,{p}}$ with $\Lambda_{p-1,{p}}^p$ in the above definition and equation (\ref{ 23}).  
It is clear  that 
 the corresponding matrix ${\mathcal A}$ proves to be singular. Under some assumption this equation is equivalent to a system of differential equations in Cauchy form and algebraic ones. The algebraic equations can be treated as constraints on the multibody system motion. It means that we may introduce `new' coordinates, {\em e.g.}, Euler angles or Fedorov vector--parameter, in order to obtain the  Lagrange equation with a non--singular symmetric matrix ${\mathcal A}$.\medskip
 
\underline{\em  Scholium}.  The singularity of Lagrange description in different generalized coordinates and velocities is the price that we must pay if we give up Newton-Euler description in kinematical twists. In practice this price is not very high.

\medskip 

\section{A continuum} Suppose the set $A \in {\bf X}_3$ has no pure point of the measure $\mu_{_{LS}} $, $\mu_{_{LS}} (dx)=\mu _3(dx)$ and $m(dx)=\rho _x\mu _3(dx)$ in $A^c$.\medskip

\noindent{\bf Strain matrix and its rate}. 
Given $x(t)$ and $y(t)\in A^c$ in the instant $t\in {\bf T}$, 
 define their radius--vectors $\vec r_x(t)$ and $\vec r_y(t)$ (in ${\cal E}_0$) and  the vector $\vec h (t)=\vec r_y-\vec r_x (t)$. 
If $\vec h (t)$ is small we have 
\[
v^0_y(t)\cong v^0_{x}(t)+dv^0_{x}/d(r_x^0)h^0(t)
\]
Define the matrix $Z_x(t)$ as the solution of the following equation 
\[
Z_x^{\centerdot }(t)=dv^0_{x}/d(r_x^0)
\]
with initial data $Z_x=I$ for $t=t_0$.
\begin{Definition} {\em \cite{Konoplev1999}} The matrices $Z_x$ and $Z_x^{\centerdot }$ are called {\em strain}  one and its {\em rate} at the point $x(t)\in A^c$ in the instant $t$, respectively.
\end{Definition}
There is no reason to consider the strain matrix and its {rate} as important (kinematical) characteristics of continuum motion. \medskip%

\noindent{\bf Stress matrix}. Let us define (see also  {\cite{Konoplev1996,Konoplev1999}}): \vspace{-3pt}
\begin{enumerate}{\parskip -0.15cm 
\item  a section $S$  between the set $A$ and an arbitrary plane $P$;
\item  the  vector bi--measure 
\[
{\mathcal D}(A) =\int \hspace{-.05cm} \chi _{_A} l^{\Delta(x,x^e)} \mu _3(dx)\]
where elements of the slider function $l^\Delta $ are axial at $x\in A$;
\item  on the set $S$ the slider function $l
^{\delta (x)}$ of the measure $%
{\mathcal 
D}$ w.r.t. Lebesgue
$2-$dimensional measure $\mu _2(dx)$ on Borel $\sigma $--algebra $\sigma
_2$ of open subsets of $S$ such that 
\[
{\mathcal D}(S) =\int \hspace{-.05cm} \chi _{_S} l^{\delta(x,x^e)} \mu _2(dx)
\]
\item  $3\times 3$--matrix--function $T_x$ of $x$ and $t$ which can be differentiable by $%
x$ the necessary number of times and such that the vector $\vec {\delta}(x,x^e)$ has the coordinate representation 
\[{\delta}^0(x,x^e)
=T_xn^0_x\]
where ${\vec n_x}$ is the normal to the plane $P$ at the point $x$;
\item  the entries of ${\Delta }^0(x,x^e)$ being
connected with the rows 
$T_{x}^{j}$ ($j=\overline{1,3}$) of the matrix $T_x$ by the following relation (in the frame ${\cal E}_0$)}
\[
\Delta ^0(x,x^e)={\rm   Div}{\it T_x}\stackrel{\rm { def}}{=}{\rm  col}\{{\rm   div}%
{\it T}_{\it x}^{1},{\rm   div}{\it T}_{\it x}^{2},{\rm   div}%
{\it T}_{\it x}^{3}\}
\]
\end{enumerate}
\begin{Remark}   One may see that the measure ${\mathcal D}(A)$ is introduced under the influence of Gauss--Ostrogradsky divergence theorem  \cite{Truesdell}, but here it is
said nothing about the  properties of $T_x$, {\em e.g.}, about its symmetry.\end{Remark}%

\begin{Definition}
 {\rm \cite{Konoplev1999}}\ 1.  The slider function $l^{\Delta (x,x^e)}$ is called {\em intensity of stress action} upon\newline \phantom{pii} \hspace{ .38cm} $x\in A^c$;
 
\hspace{ .38cm} 2. $T_x$ is called {\em stress matrix}.\end{Definition}

\noindent{\bf Notion of continuum}. 
A matrix--function of entries of some matrices 
is called {\em isotropic} if it is invariant w.r.t. ${\mathcal SO}({\bf R},3)$. Assume that the matrices ${T}_{x}$, $Z_x$ and $Z_x^{\centerdot }$ are  invariant w.r.t. rotations. An isotropic map ${T}_{x}(Z_x, Z_x^{\centerdot })$ is called {\em constitutive} or {\em stress--strain} relation.

Let us note the set of all isotropic  maps from the strain matrix $Z_x$ and its rate $Z_x^{\centerdot }$ to the stress matrix $T_x$  as $\alpha^2 (Z_x, Z_x^{\centerdot })$.
\begin{Definition}
Suppose that 
\begin{equation}
l^{\phi(x,x^e)}=\rho_xl^{g (x,x^e)},\quad l^{c(x,x^e)}=l^{\Delta (x,x^e)} \label{ 6}
\end{equation}
the stress matrix $T_x$ belongs to $\alpha^2 (Z_x, Z_x^{\centerdot })$
and the measure of inertia is time--constant on $A^c$, {\em i.e.}, $\frac d{dt}m(dx)=0$. Then the set $A$ is called {\em continuous medium} or {\em continuum}.
\end{Definition}

\noindent{\bf Motion of continuum}. 
Due to relations (\ref{ 13a}) and (\ref{ 6}) the {equation of continuum motion} at a
point $x\in A^c$ is of the form (see also \cite{Konoplev1999,Truesdell}) (in the
Galilean frame ${\cal E}_0$) 
\begin{equation}
	\rho _xv^{0\centerdot}_{x}=\rho _x g^0+{\rm  Div}
{\it T}_{\it x}, 
\quad \rho _x^{\centerdot}+{\rm   div}\rho _x \hskip .05cm v^0_x=0, \quad T_x=T_x(Z_x,Z_x^{\centerdot }) \in \alpha ^2(Z_x,Z_x^{\centerdot })
\label{mot}
\end{equation}
where ${\rm  Div}{\it T}_{\it x}$ is the constraint action \cite {K}.\medskip

\noindent{\bf Some constitutive relations}. Constitutive relations define the properties of a continuum and its motion equation.\medskip

\underline{\em $3-$dimensional case}.
 For any $3\times 3$--matrix $A$ the
aggregate $PAQ$ is an isotropic function of $A$ if the matrices $P$ and $Q$ are
proportional to $I$ with scalar coefficients being invariant w.r.t. rotations.  

Define the following linear combination 
\begin{equation}
T=r_1{E}_1+{\it r}_2{E}_2+{\it r}_3{E}_3
\label{ 7}
\end{equation}
where $r_i$ are invariant w.r.t. rotations (they can be functions of the
time, invariants of  $U_x$ and so on), $U_x=Z_x$ or $U_x=Z_x^{\centerdot }$;
 \begin{equation}
E_1=({\rm   tr }{\it U_x})I,\ E_2={\rm   sym}{\it U_x}=0.5({\it U_x}+{\it U_x^{T}}),\ E_3={\rm   ant}{\it U_x}=0.5({\it U_x}-{\it U_x^{T}})\label{ 88}
\end{equation}
\begin{Theorem} {\em \cite{Dubrovin}} All isotropic $3\times 3$--matrix functions of entries of $U_x$ are given by relation  (\ref{ 7}).
\end{Theorem}
Thus due to the theorem we may define the  following relation
\begin{equation}
T_x=-r_0 E_0 +r_1{E}_1+r_2{E}_2+r_3{E}_3, \quad E_0=I
\label{ 8}
\end{equation}
as the most general linear {\em constitutive} one. It is conventional the invariant w.r.t. rotations $r_i$ to be called {\em rheological
coefficients} (w.r.t. the set of $E_i$). The constitutive relation 
(\ref{ 8}) is called {\em quasi--linear} if its  rheological
coefficients are functions of $U_x$--matrix invariants (in particular it means that the summand  $r_1E_1$ could be omitted in quasi--linear constitutive relations).

If $U_x=Z_x$ and $r_0=0$ the continuum is called {\em elastic material}, if $%
U_x=Z_x^{\centerdot }$ and $r_0>0$ (called Pascal pressure) the continuum
is called {\em viscous fluid} \cite{Lur'e}.
\begin{Remark}  Continua defined by relations (\ref{ 88})--(\ref{ 8}) coincide with the continua
used in continuum mechanics in the following cases {\em \cite{Truesdell,Lur'e}}\vspace{-7pt}
\begin{enumerate}{\parskip -0.1cm 
\item the Pascal pressure $r_0$ is positive  and $r_1=r_2=r_3=0$ ({%
ideal fluid});
\item $r_0$ is non--negative and $r_3=0$ ({continua of
Navier--Stokes--Lame type});
\item $r_0$ is non--negative and $r_1{\rm   tr }{\it I}+{\it r}_2=0\rightarrow 
{\rm   tr }{\it T_x}=-{\it r}_0{\rm   tr }{\it I}$ (continua used in some theories).}
\end{enumerate}
\end{Remark}

In order to use relations (\ref{ 88})--(\ref{ 8}) in the motion equation (\ref{mot}) we must calculate \vspace{-3pt}
\begin{eqnarray}
{\rm Div\hspace{0.03cm}}T_x   &=&  -{\rm \hspace{0.01cm}grad\hspace{0.05cm}} r_0+{\rm tr}\hspace{0.05cm}U_x\hspace{0.05cm}{\rm \hspace{0.01cm}grad\hspace{0.05cm}}r_1+r_1{\rm \hspace{0.01cm}}{\rm Div\hspace{0.05cm}}({\rm tr}\hspace{0.05cm}U_x\hspace{0.05cm}I)+ 0.5r_2({\rm Div\hspace{0.03cm}}{\it U_x}+{\rm Div\hspace{0.03cm}}{\it U_x^{T}}) +\nonumber \\
&&0.5r_3({\rm Div\hspace{0.03cm}}{\it U_x}-{\rm Div\hspace{0.03cm}}{\it U_x^{T}})+0.5({\it U_x}+{\it U_x^{T}}){\rm \hspace{0.01cm}grad\hspace{0.05cm}}r_2 + 0.5({\it U_x}-{\it U_x^{T}}){\rm \hspace{0.01cm}grad\hspace{0.05cm}}r_3 
\label{2_ 36}\vspace{-7pt}
\end{eqnarray}
 Introduce the next notations 
\[ U_x=
\begin{bmatrix}
u_1^1\ & \  u_2^1 & \ \ u_3^1 \\ 
u_1^2\ & \  u_2^2 & \ \ u_3^2 \\ 
u_1^3\ & \  u_2^3 & \ \ u_3^3
\end{bmatrix},\quad 
u_{ik}^j=\frac {\partial u_i^j}{\partial r_k^0},\quad r_x^0={\rm col}(r_1^0,r_2^0,r_3^0)  
\]
Let $u^{j}$ be the rows of $U_x$, then with the help of routine calculations   we have 
${\rm Div\hspace{0.05cm}} ({\rm tr}\hspace{0.05cm}U_x\hspace{0.05cm}I) ={\rm col\hspace{0.05cm}}\{ u_{11}^1,u_{22}^2,u_{33}^3\}$ and \vspace{-3pt}
\[
{\rm Div}\hspace{0.03cm}U_x=
\begin{pmatrix}
u_{11}^1+u_{22}^1+u_{33}^1 \\ 
u_{11}^2+u_{22}^2+u_{33}^2 \\ 
u_{11}^3+u_{22}^3+u_{33}^3
\end{pmatrix},\quad  
{\rm Div\hspace{0.03cm}}U_x^{T}= 
\begin{pmatrix}
u_{11}^1+u_{12}^2+u_{13}^3 \\ 
u_{21}^1+u_{22}^2+u_{23}^3 \\ 
u_{31}^1+u_{32}^2+u_{33}^3
\end{pmatrix} 
\]\medskip

\underline{\em $2-$dimensional case}.
In the case of $2\times 2$--matrices it is easy to see that for
matrices $P$ and $Q$ the aggregate $PU_xQ$ is an isotropic  map of $U_x$ if $P$ and $Q$ are
of the kind $aI+\widetilde{a}\widetilde{I}$ where $\widetilde{I}=
\begin{bmatrix}
0 & -1 \\ 
1 & 0
\end{bmatrix} $, the scalar coefficients $a$ and $\widetilde{a}$ are invariant w.r.t. rotations. 

Introduce the following matrices  $E_2={U_x}$, 
$E_3={U_x}^T$, $E_4=\widetilde{I}{U_x}$, 
$E_5=\widetilde{I}{U_x}^T$, 
$E_6={U_x}\widetilde{I}$, 
$E_7={U_x}^T\widetilde{I}$, 
$E_8=\widetilde{I}{U_x}\widetilde{I}$, 
and $E_9=\widetilde{I}{U_x}^T\widetilde{I}$. The  linear combinations of the matrices $E_i$ ($i=\overline{2,9}$) generate a manifold with a basis consisting of the linear independent matrices $E_2$, $E_3$, $E_4$ and $E_5$. 

We may use this basis in order to define the following relation
\begin{equation}
T_x = - r_0 I+\widetilde{r_0}\widetilde{I}+r_1{\rm   
tr}{\it U} I+\widetilde{r_1}{\rm   
pf}{\it U}\widetilde{I}+ r_2{\it U_x}+r_3U^T_x 
+r_4\widetilde{I}U_x+r_5\widetilde{I}U_x
\widetilde{I},\ {\rm   
pf}{\it U}={\rm   tr }\{{\it \widetilde{I}U}\} \label{ 10}
\end{equation}
as {\em constitutive relation} with rheological
coefficients $r_i$ and $\widetilde{r_i}$ (w.r.t. the set of $E_i$)
being invariant w.r.t. rotations. \medskip

\underline{\em Correct continua}. 
A continuum  is called {\em correct} if the corresponding constitutive relation is invertible \cite{Konoplev1999} (see map (\ref{ 8}) and (\ref{ 10})).

Let us stop at $2-$ and $3-$dimensional constitutive relations with the same
`structure'. With the help of routine calculations we see the following statement to be true.
\begin{Theorem} 
In $2$-- and $3$--dimensional cases let constitutive relations be of the 
(\ref{ 8})--form and $(r_1{\rm   tr }{\it I}+{\it r}_2){\it r}_2{\it r}_3\neq 0$ where $I$ is used as $2$-- and $3$--dimensional identity matrices, respectively. Then there
exists the inverse map 
\[
U_x=n_0I+n_1({\rm   tr }{\it T_x)I}+{\it n}_2\hspace{0.05cm}{\rm   sym}{\it T_x}+{\it n}_3\hspace{0.05cm}{\rm   ant}%
{\it T_x}
\]
where
\[
n_0 =\frac{r_0}{r_1{\rm   tr }{\it I}+{\it r}_2},\ n_1=\frac{-r_1}{r_2(r_1{\rm   %
tr}{\it I}+{\it r}_2)}, \ n_2 =\frac 1{r_2},\ n_3=\frac 1{r_3}
\]
\end{Theorem}
Thus  Navier--Stokes--Lame continua are
incorrect as $r_3=0$.\medskip

\underline{\em Rheological coefficients and moduli (ratio)}. The three coefficients 
\[
\varepsilon=\frac 1{n_2-n_1}=\frac{r_2(r_1{\rm   tr }{\it I}+{\it r}_2)}{r_1(%
{\rm   tr }{\it I}-1)+{\it r}_2},\ { \mu }=\frac 1{2n_2}=\frac {r_2}{2},\
{ \nu } =\frac{n_1}{n_1-n_2}=\frac{r_1}{r_1({\rm   tr }{\it I}-1)+{\it r}_2}
\]
can be called {\em Young modulus}, $\varepsilon$, {\em shear or rigidity one}, ${\mu }
$, and {\em Poisson ratio}, ${ \nu }$, respectively (see also {\rm \cite{Konoplev1999}}). Note that there is the known relation $\varepsilon =2{\mathcal \mu }(1+{\mathcal \nu })$. 

In this way of definition 
Young modulus, shear or rigidity one and  Poisson ratio depend on continuum dimensions.

Let us define now 
 \begin{equation}E_0=I, \quad 
	{E}_1=({\rm   tr }{\it U_x)I},\quad {{\it E}}
_2={\it U_x},\quad {{\it E}}_3={\it U_x^{T} }\label{du}
\end{equation}
then  we may take the constitutive relation (\ref{ 8})
with new rheological coefficients ${r}_0$, ${r}_1$, ${r}_2$ and $%
{r}_3$ (w.r.t.  the new set (\ref{du})) and new Young modulus, shear or rigidity one and  Poisson ratio. Thus they are not unique. \medskip

\section{Brief comments}
Continuum mechanics is closely connected with Riemann integral theory. In continuum mechanics as well as in Riemann theory there is realized the idea of approximating an area by summing rectangular strips (segments, squares or boxes), then using some kind of limit process to obtain the exact area required. It is safe to say that we may name the well known mechanics of continua as that of Cauchy (due to the man who created it).

The Riemann integral, natural though it is, has been superseded by the Lebesgue or Lebesgue--Stieltjes integral and other more recent theories of integration. 
In this way, V. Konoplev suggested a new architecture of continuum mechanics based on Lebesgue integral and his algebraic theory of screws. As result in Konoplev mechanics there do not arise boxes or particles which can be rotated by the laws of Newtonian mechanics as well as there are no imaged surfaces with stresses over them and other concepts of Cauchy mechanics. But with introducing the measures as Lebesgue integrals he was forced to exclude mass--points and their systems from consideration.
 
Unlike V. Konoplev, we use Lebesgue--Stieltjes integral 
 in order to introduce main mechanics measures and classes of mechanical systems such that mass--points, rigid bodies and  continua (under the special assumption about interaction in mechanical systems and their `constitution'). In this way we  become  closer  to mechanics of C. Truesdell.

\section*{Conclusion} 
It is a first attempt to represent elements of Konoplev's axiomatics and its (possibly debatable) modification in the form of a journal paper. 
One must realize the difficulties and gaps issued from this goal. 

It is impossible to separate the theory given above from that of Konoplev. That is why the paper author prefers to yield the palm to Prof. V. Konoplev but carries full responsibility for all lacks of this paper. This is the place to express his sincere thanks to Prof. V. Konoplev
for the collaboration of many years.

\newpage
\subsubsection*{\bf Appendix: A review}

{\em Dear Professor Cheremensky},\medskip

I have looked at your paper, and I find it nearly incomprehensible. I regret to say that I cannot consider it for the Archive of Rational Mechanics, as it does not meet basic standards of clarity that would allow me to review it.
I should add that, even if it could be rewritten in a comprehensible way, the subject matter is probably not suitable for the Archive today.  During a brief period in the 1970s, the Archive did become a forum for some axiomatic work in mechanics, but in hindsight this has become some of the least influential work that has been published in the Archive, and we do not encourage it now.  Generally, we also encourage authors to use the most conventional notation and to avoid abstraction for the sake of abstraction, so that the work is readable by the widest possible scientific audience.
Thank you for your interest in the Archive and I hope you find a receptive audience for your work.
\newline {\em \phantom{jn} 
\hspace{2cm} Sincerely yours, \newline \phantom{jn} \hspace{9cm} Richard James, 
Editor in Chief}\bigskip

The works of C. Truesdell (Founder and Editor of Archive for Rational Mechanics and Analysis: 1952--1989), W. Noll and B.D. Coleman, {\em etc.}, are published in this journal.  We are not assured  that they are really {\em least influential}. They are contributed to foundational rational mechanics, whose aim is to construct a mathematical model for treating (continuous) mechanical phenomena. 


We would be highly grateful with whoever would bring any element likely to be able to make progress the development, and thus the comprehension, of the paper. Any comments, reviews, critiques, or objections are kindly invited to be sent to the author 
 by e--mail.
\end{document}